\documentclass[superscriptaddress,showpacs,amssymb,10pt,reprint,aps,prd,longbibliography,nofootinbib,floatfix]{revtex4-2}

\usepackage{graphicx,epsfig,amssymb,times} 
\usepackage{amsmath,amsfonts,amsthm}
\usepackage{bm}
\usepackage{epstopdf}
\usepackage{hyperref}
\usepackage[caption=false]{subfig}
\usepackage[usenames]{color}   
\usepackage[dvipsnames]{xcolor}
\usepackage[normalem]{ulem}
\usepackage{multirow}
\usepackage{orcidlink}

\definecolor{coolblack}{rgb}{0.0, 0.18, 0.39}
\definecolor{darkred}{rgb}{0.5,0,0}
\definecolor{darkgreen}{rgb}{0,0.5,0}
\definecolor{darkblue}{rgb}{0,0,0.5}
\definecolor{lapislazuli}{rgb}{0.15, 0.38, 0.61}
\definecolor{venetianred}{rgb}{0.78, 0.03, 0.08}
\definecolor{bleudefrance}{rgb}{0.19, 0.55, 0.91}
\definecolor{dogwoodrose}{rgb}{0.84, 0.09, 0.41}
\hypersetup{colorlinks=true, citecolor=darkblue, linkcolor=darkblue, 
urlcolor = darkblue}

\DeclareMathOperator{\sech}{sech}

\begin{document}

\title{\large Probing nonlinear electrodynamics-sourced black holes via light and orbital mechanics}

\author{Marco A. A. de Paula \orcidlink{0000-0002-9808-7495}}
\email{marcodepaula@ufpa.br}
\affiliation{Faculdade de Ciências Naturais, Universidade Federal do Par\'a, Campus Universit\'ario do Tocantins-Cametá, 68400-000, Camet\'a, Par\'a, Brazil.}
\affiliation{Departamento de F\'isica, Universidade Federal da Para\'iba, 58051-970, Jo\~ao Pessoa, Para\'iba, Brazil.}
	
\author{Ednaldo L. B. Junior \orcidlink{0000-0001-7230-3666}}
\email{ednaldobarrosjr@gmail.com}
\affiliation{Faculdade de Física, Universidade Federal do Pará, Campus Universitário de Tucuruí, 68464-000, Tucuruí, Pará, Brazil.}
\affiliation{Programa de Pós-Graduação em Física, Universidade Federal do Sul e Sudeste do Pará, 68500-000, Marabá, Pará, Brazil.}

\begin{abstract}

Well-founded models of nonlinear electrodynamics (NED) are typically characterized by an additional parameter associated with their deviation from linear electrodynamics. In addition to that, from a phenomenological point of view, one of the most intriguing results of NED theory is that photons follow null geodesics according to an effective geometry. In particular, for a two-parameter electromagnetic Lagrangian density, we might observe the vacuum birefringence phenomenon, i.e., photons can follow null geodesics according to two effective geometries. Over the past few decades, efforts have been made to better understand the role of effective light cones in NED theory, as well as to constrain the additional free parameters provided by the theory. We investigate the ModMax black hole (BH) geometry, which is a well-motivated NED-sourced BH based on a two-parameter Lagrangian density, via light and orbital mechanics, considering the Shapiro time, the Sagnac effect, the gravitational and kinematic shifts, and the orbital precession. We find that the corrections introduced by nonlinear electromagnetic fields can be significant. In particular, we show that the Shapiro time delay and the gravitational redshift are identical for both effective metrics, making these observables insensitive to the birefringent structure of ModMax electrodynamics. By contrast, the Sagnac effect and the kinematic shifts remain capable of distinguishing the two effective geometries. Moreover, the orbital mechanics investigation, considering the orbital periapsis precession of the S2 star around Sagittarius A$^{\star}$, provides an observational constraint on the coupling between the NED parameter and the charge-to-mass ratio of the ModMax BH. Generally speaking, we provide a step-by-step procedure for theoretically exploring signatures of NED fields and constraining the free parameters of NED-sourced BH models.

\end{abstract}

\date{\today}

\maketitle

\section{Introduction}

The acceptance of black holes (BHs) as part of physical reality is grounded in the systematic experimental validation of general relativity (GR) in both weak- and strong-field regimes. For example, in the weak-field regime, we have the perihelion precession of Mercury's orbit~\cite{Einstein:1916vd,Park:2017zgd}, the deflection of light by the Sun~\cite{Dyson:1920cwa,Crispino:2019yew}, and the gravitational redshift of light~\cite{Pound:1964zz,Pound:1965zz,Vessot:1980zz}. Conversely, in the strong-field regime, we have a collection of gravitational wave detections~\cite{LIGOScientific:2016aoc,cardiff_gw_catalog,gwosc_gwtc} and the shadow images of supermassive BHs at the centers of galaxies~\cite{EventHorizonTelescope:2019dse,EventHorizonTelescope:2019ggy,EventHorizonTelescope:2022wkp,EventHorizonTelescope:2022xqj}. Some of these tests work both ways: just as they help us better understand the theory of gravity used to describe the central objects, they also help us better understand the properties of these central objects. This allows us to use the astrophysical data associated with the experimental validation of GR to better understand the physical feasibility of alternative theories of gravity, as well as alternative BH solutions\footnote{By ``alternative BH solutions,'' we mean those other than the standard BH solutions of GR, namely, Schwarzschild, Reissner–Nordström (RN), Kerr, and Kerr-Newman BH solutions.}.

Despite its experimental robustness, GR has some problems that lead one to believe it is not the complete theory of gravity. Among these problems, we highlight the prediction of curvature singularities at the center of the standard BH solutions~\cite{wald2010general,Hawking:1973uf} and the inability to adequately explain the earlier stages of the universe~\cite{Hawking:1970zqf} as well as the last stages of BH evaporation~\cite{Hawking:1975vcx}. To circumvent some of these problems, we can explore alternative theories of gravity by modifying the gravitational sector of standard GR (see, e.g., Refs.~\cite{Shankaranarayanan:2022wbx,lobo2008dark} and references therein). Alternatively, it is also possible to modify the energy-matter sector of standard GR by considering energy-momentum tensors that differ from those used to obtain the standard BH solutions in GR. In particular, by considering nonlinear electrodynamics (NED) models. 

The NED framework was originally proposed in the 1930s as a completion of linear electrodynamics for sufficiently strong electromagnetic fields. This was the case, e.g., with the Born-Infeld~\cite{born1934foundations,born1934quantum} and Euler-Heisenberg~\cite{Euler:1935qgl,Heisenberg:1936nmg} electrodynamics. In the late 1990s, it was discovered that NED models can be used to obtain regular BH geometries~\cite{Ayon-Beato:1998hmi}, i.e., BHs free of curvature singularities. Since then, the NED framework has been widely used to obtain an extensive list of BH solutions in GR and beyond (see, e.g., Refs.~\cite{Lan:2023cvz,Bronnikov:2022ofk,Junior:2015fya,deSousaSilva:2018kkt} and references therein). For reviews on NED and its applications, we recommend Refs.~\cite{Sorokin:2021tge,Delphenich:2003yw,Fouche:2016qqj}.  We emphasize, however, the phenomenological appeal of nonlinear electromagnetic fields. According to the NED framework, electromagnetic waves satisfy an eikonal equation that is different from that of linear electrodynamics~\cite{Gutierrez:1981ed,Novello:1999pg}. This has implications for the motion of photons, leading to signatures of NED fields in null geodesics~\cite{Vrba:2019vqh,Habibina:2020msd,Amaro:2020xro,Habibina:2021iuq,Breton:2021mju,Toshmatov:2021fgm,Ryotaky:2023}, gravitational lensing~\cite{Eiroa:2005ag,Liang:2017vdd,Liang:2017wym,Ghaffarnejad:2016dlw,2019ApJ87412S,dePaula:2023ozi,dePaula:2026blu}, and quasinormal modes~\cite{Moreno:2002gg,Nomura:2020tpc,Nomura:2021efi,Toshmatov:2018ell,Toshmatov:2018tyo} of NED-sourced BH spacetimes. The causality of such spacetimes is also affected~\cite{Russo:2024llm,Murk:2024nod,dePaula:2024yzy,Schellstede:2016zue}, as well as their thermodynamic properties~\cite{Abe:2025vdj,Bokulic:2021dtz}.

Interestingly, the analysis of some tests of GR, considering NED-sourced spacetimes as seeds, has somehow been overlooked in the literature. Yet, this is important because it can help us better understand the properties of these objects. Moreover, some well-motivated NED-based BH models have free parameters that can be constrained based on the experiments used to validate GR. Among the potentially relevant models, we consider the ModMax BH~\cite{Flores-Alfonso:2020euz}. This model is interesting because it is invariant under both duality rotations and conformal transformations~\cite{Kosyakov:2020wxv,Bandos:2020jsw}, preserving the same symmetries as Maxwell's theory. Additionally, it provides a BH configuration with a clear physical interpretation: the nonlinear parameter acts as a screening factor for the BH charge in the metric function, which keeps the line element simple. Furthermore, as the ModMax BH model is singular\footnote{Not every well-motivated NED-based BH spacetime is regular. A notorious example, besides the ModMax case, is the Euler-Heisenberg BH~\cite{Breton:2021mju,dePaula:2026blu}.} and obtained from a two-parameter Lagrangian density, it is not prone to the Laplacian instability pointed out in Ref.~\cite{DeFelice:2024seu} that occurs for a broad class of NED-based BH geometries.

We aim to provide a step-by-step general procedure for theoretically exploring the signatures of nonlinear electromagnetic fields and constraining the free parameters of the corresponding NED-sourced BH models in light of some GR experimental tests. To do that, we restrict our analysis to the Shapiro time, Sagnac effect, gravitational and kinematic shifts, and orbital precession. The Shapiro time delay, Sagnac effect, and shifts are related to the trajectories of light, allowing us to explore the effective geometry followed by photons in NED. Simply put, the Shapiro time is the delay of a light signal when propagating from Earth to another celestial body and reflecting back to Earth~\cite{Shapiro:1964uw,Shapiro:1968zza}. In turn, the Sagnac effect is the proper time difference between two light rays moving in opposite directions along closed paths~\cite{sagnac1913ether,sagnac1913preuve}. Moreover, roughly speaking, the gravitational and kinematic shifts correspond to the radiation emitted by a star in a circular orbit, which can be measured by a detector~\cite{hobson2006general}. These effects have been widely used to probe alternative theories of gravity and matter distributions~\cite{Junior:2023nku,Bhattacharya:2025qcd,Hsieh:2021rru,Ben-salem:2022txj,Hook:2017psm,Poddar:2021sbc,Hou:2017cjy,DellaMonica:2023ydm,Fathi:2019jgd,Ashtekar:1975wt,Souza:2024ltj,Moradpour:2026ppq,Aliberti:2024udm,martinez2024observational,MosqueraCuesta:2004em,Guzman-Herrera:2023zsv}. Finally, the orbital precession is the gradual, slow rotation or shift in the orientation of an orbiting body's elliptical path over time~\cite{d2022introducing,hobson2006general}. Recently, it has been used to constrain the physical parameters of BH solutions in GR and beyond~\cite{Yang:2026glg,Yang:2024fgm,Yang:2024xqa,QiQi:2024dwc,Lin:2023rmo}, particularly using the data for the orbital precession of the S2 star around Sagittarius A$^{\star}$ (Sgr A$^{\star}$), as measured by the GRAVITY Collaboration~\cite{GRAVITY:2020gka}.

Our results indicate that nonlinear electromagnetic fields strongly affect the Sagnac effect and the kinematic shifts, but the Shapiro effect and the gravitational redshift alone cannot be used to discriminate between the effective geometries followed by photons in NED.\footnote{The gravitational and kinematic shifts considering the ModMax BH were investigated in Ref.~\cite{Guzman-Herrera:2023zsv} using an unconventional definition for the energy and angular momentum of photons in NED-sourced BH spacetimes. Here, we present our results using the standard definition found in the literature (see, e.g., Refs.~\cite{dePaula:2023ozi,2019ApJ87412S} and references therein). In Appendix~\ref{breton}, we explain in detail how our results differ from those of Ref.~\cite{Guzman-Herrera:2023zsv}.} Moreover, the analysis of orbital mechanics provides a physical constraint on the possible values of the coupling between the NED parameter and the charge-to-mass ratio of the corresponding BH in the ModMax theory. The remainder of this work is organized as follows. In Sec.~\ref{sec:setup}, we present the NED framework for a two-parameter Lagrangian density, considering the general framework (\ref{subsection:gf}) and the ModMax electrodynamics (\ref{subsec:me}). The ModMax BH geometry is reviewed in Sec.~\ref{sec:bhsmt}. In Sec.~\ref{Shap}, we introduce the Shapiro time delay. We first present the general formalism (\ref{subsec:gf}) and the time delay considering the ModMax (\ref{subsec:tdem}). We then consider specific benchmark values for the BH parameters to perform a numerical analysis (\ref{subsec:natt}). In Sec.~\ref{sec:se}, we investigate the Sagnac effect. For simplicity, we provide a pedagogical introduction (\ref{subsec:preli}), the general equations in NED (\ref{subsec:sened}), and then discuss the Sagnac effect in the ModMax BH spacetime (\ref{subsec:semg}). The gravitational and kinematic redshifts are investigated in Sec.~\ref{sec:gkr}. We first introduce the main equations (\ref{subsec:framework}), and then we present the main results (\ref{subsec:mr}). The massive particle effects (orbital precession) are presented in Sec.~\ref{sec:mpe}. Our final remarks are presented in Sec.~\ref{sec:cr}. Here, we consider the signature $+2$ ($-, +, +, +$) and use the natural units for which $G = c = \hbar = 1$, unless otherwise stated. 

\section{Setup}\label{sec:setup}

In this section, we present the field system associated with a two-parameter NED Lagrangian minimally coupled to GR. We then introduce the ModMax electrodynamics, which is used to obtain the ModMax BH. We also present the effective geometries followed by photons within this framework.

\subsection{General framework}\label{subsection:gf}

The action that governs the minimal coupling between GR and NED can be written as~\cite{Allahyari:2019jqz}
\begin{equation}
\label{S}\mathrm{S} = \dfrac{1}{16\pi}\int \sqrt{-g}\left[R-\mathcal{L}(F,G) \right] d^{4}x,
\end{equation}
where $g = \det(g_{\mu\nu})$, with $g_{\mu\nu}$ being the covariant components of the metric tensor \textbf{g}, $R$ is the Ricci scalar, and $\mathcal{L}(F,G)$ is a gauge-invariant electromagnetic Lagrangian density. The electromagnetic scalar invariants $F$ and $G$ are defined by
\begin{equation}
\label{maxwellscalar}F \equiv F_{\mu\nu}F^{\mu\nu} \ \ \text{and} \ \ G \equiv F_{\mu\nu}\star F^{\mu\nu},
\end{equation}
respectively. The Faraday tensor $F_{\mu\nu}$ is given by
\begin{align}
\label{Fmunu}F_{\mu\nu} = \nabla_{\mu}A_{\nu} - \nabla_{\nu}A_{\mu},
\end{align}
in which $A_{\mu}$ is the four-vector potential and $\star F^{\mu\nu}$ is the dual electromagnetic field tensor, namely,
\begin{equation}
\label{dual}\star F^{\mu\nu} = \dfrac{1}{2}\epsilon^{\mu\nu\sigma\rho}F_{\sigma \rho}.
\end{equation}
The symbol $\star$ represents the Hodge dual operator, and $\epsilon_{\mu\nu\sigma\rho}$ denotes the Levi-Civita tensor, which satisfies $\epsilon_{\mu\nu\sigma\rho}\epsilon^{\mu\nu\sigma\rho} = -4!$. Moreover, the Faraday tensor obeys
\begin{equation}
\label{EFTC1}\nabla_{\mu}\left(\mathcal{L}_{F}F^{\mu\nu} + \mathcal{L}_{G}\star F^{\mu\nu}\right)  = 0,
\end{equation}  
where $\mathcal{L}_{F} \equiv \partial\mathcal{L}/\partial F$ and $\mathcal{L}_{G} \equiv \partial\mathcal{L}/\partial G$, and
\begin{equation}
\label{EFTC2}\nabla_{\mu}\star F^{\mu\nu} = 0.
\end{equation}

The Einstein-NED (ENED) field equations are derived by varying the action~\eqref{S} with respect to $g^{\mu\nu}$. Thus, we get
\begin{equation}
\label{E-NED_F}G_{\mu}^{\ \ \nu} = 8\pi T_{\mu}^{\ \ \nu} = 2\mathcal{L}_{F}F_{\mu\alpha}F^{\nu\alpha}+\dfrac{1}{2}\left(G\mathcal{L}_{G}-\mathcal{L}\right)\delta_{\mu}^{\ \ \nu}.
\end{equation}
By restricting our analysis to configurations where $G = 0$, the correspondence with standard linear electrodynamics (Maxwell's theory) in the weak-field limit is recovered when 
\begin{equation}
\mathcal{L}(F) \rightarrow F \quad \text{and}  \quad \mathcal{L}_{F} \rightarrow 1,
\end{equation}
for $F \ll 1$. Consequently, the setup described above reduces to the standard Einstein-Maxwell theory in this limit~\cite{wald2010general}.

In NED, photons follow null geodesics in the background of an effective metric $\bar{g}^{\mu\nu}$, which, in general, differs from the spacetime metric $g^{\mu\nu}$~\cite{plebański1970lectures,Gutierrez:1981ed,Novello:1999pg}. For two-parameter Lagrangian densities, the NED models can exhibit the birefringence phenomenon. In this case, light rays can propagate along two or more effective light cones according to their polarizations. The effective metrics can be written as~\cite{Allahyari:2019jqz}
\begin{align}
\nonumber \bar{g}^{\mu\nu}_{\pm} = \ & \mathcal{L}_{F}g^{\mu\nu}-4\big[\left(\mathcal{L}_{FF}+\Omega_{\pm}\mathcal{L}_{FG} \right)F^{\mu}_{\ \ \lambda}F^{\lambda\nu}+ \\
\label{eff_metric}& \left(\mathcal{L}_{FG}+\Omega_{\pm}\mathcal{L}_{GG}\right)F^{\mu}_{\ \ \lambda}\star F^{\lambda\nu}\big],
\end{align}
where
\begin{equation}
\label{Omega}\Omega_{\pm} = \dfrac{-\Omega_{2}\pm\sqrt{\left(\Omega_{2}\right)^{2}-4\Omega_{1}\Omega_{3}}}{2\Omega_{1}}.
\end{equation}
The functions $\Omega_{1}$, $\Omega_{2}$, and $\Omega_{3}$ are given by
\begin{align}
\Omega_{1} =\  & \mathcal{L}_{FG}\left(2F\mathcal{L}_{GG}-\mathcal{L}_{F}\right) + G\left(\mathcal{L}_{GG}^{2}-\mathcal{L}_{FG}^{2} \right), \\
\nonumber \Omega_{2} = \ & \left(\mathcal{L}_{F}+2G\mathcal{L}_{FG} \right)\left(\mathcal{L}_{GG}-\mathcal{L}_{FF} \right) + \\
& 2F\left(\mathcal{L}_{FF}\mathcal{L}_{GG}+\mathcal{L}_{FG}^{2} \right), \\
\Omega_{3} = \ & \mathcal{L}_{FG}\left(2F\mathcal{L}_{FF}+\mathcal{L}_{F}\right) + G\left(\mathcal{L}_{FG}^{2}-\mathcal{L}_{FF}^{2} \right),
\end{align}
respectively. In the remainder of this paper, we use the term ``massless particles'' (``photons'') to refer to particles that follow null geodesics in the spacetime metric $g^{\mu\nu}$ (effective metrics $\bar{g}^{\mu\nu}_{\pm}$). Moreover, the signs $\pm$ in Eq.~\eqref{eff_metric} denote the two possible paths for the propagation of light, each related to a distinct polarization of light. From now on, we refer to the polarizations governed by $\bar{g}^{\mu\nu}_{+}$ and $ \bar{g}^{\mu\nu}_{-}$ as polarizations $\mathcal{P}_+$ and $ \mathcal{P}_-$, respectively, following the same notation as in Ref.~\cite{dePaula:2026blu}.

\subsection{ModMax electrodynamics}\label{subsec:me}

The ModMax Lagrangian density can be written as~\cite{Flores-Alfonso:2020euz,Kosyakov:2020wxv}
\begin{equation}
\label{modmax}\mathcal{L}(F,G) = F \cosh\gamma-\sinh \gamma \sqrt{F^{2}+G^{2}},
\end{equation}
where $\gamma$ is the free parameter of the model. Here, we consider only $\gamma \geq 0$ because $\gamma < 0$ leads to causality issues~\cite{Bandos:2020jsw}. Notice that for $\gamma = 0$, we obtain Maxwell electrodynamics. By inserting Eq.~\eqref{modmax} into Eq.~\eqref{eff_metric}, we obtain the corresponding effective geometries in ModMax electrodynamics, namely
\begin{align}
\nonumber \bar{g}^{\mu\nu}_{\pm} = \ & \left[\cosh\gamma-\frac{1}{2} F \sinh\gamma \left(\dfrac{1}{y}+\dfrac{F \tanh\gamma}{y^{2}} \mp u \right)\right]g^{\mu\nu} \\
\label{eff_metrics} &+2 \sinh\gamma \left(\dfrac{1}{y} - \dfrac{F \tanh\gamma}{y^{2}}\pm u\right)F^{\mu}_{\ \ \lambda}F^{\lambda\nu},
\end{align}
where we defined
\begin{align}
u = u(F,G) \equiv \dfrac{\left|y-F \tanh\gamma \right|}{y^2},
\end{align}
and $y = y(F,G) \equiv \sqrt{F^{2}+G^{2}}$, for simplicity.

In the Einstein-ModMax framework, it is possible to derive BH solutions with electric or magnetic charge, as well as dyonic configurations, i.e., BH configurations with both electric and magnetic charges. However, for simplicity, we are only interested in BH solutions with electric charge. Thus, since
\begin{equation}
G = -4 \left(\vec{E}\cdot \vec{B} \right),
\end{equation}
where $\vec{E}$ and $\vec{B}$ are the electric and magnetic fields, respectively, it follows that $G = 0$ in our case. In this context, the effective metrics given by Eq.~\eqref{eff_metric} reduce to
\begin{align}
\label{eff_metric01} \bar{g}^{\mu\nu}_{+} & = \sech \gamma  g^{\mu\nu}-\dfrac{4\sinh\gamma\left(1+\tanh\gamma\right)}{F}F^{\mu}_{\ \ \lambda}F^{\lambda\nu},\\
\label{eff_metric02} \bar{g}^{\mu\nu}_{-} & = e^{\gamma} g^{\mu\nu}.
\end{align}

Recall that if two metric tensors, say $g_{\mu\nu}$ and $h_{\mu\nu}$, are conformally related, then $g_{\mu\nu} = \mathrm{A}h_{\mu\nu}$, where $\mathrm{A}$ is a non-zero differentiable function, and the null geodesics associated with the metric tensors $g_{\mu\nu}$ and $h_{\mu\nu}$ coincide~\cite{d2022introducing}. By choosing the conformal factors as $\cosh\gamma$ and $e^{-\gamma}$ for the metric functions~\eqref{eff_metric01} and~\eqref{eff_metric02}, respectively, we find that
\begin{align}
\label{eff_metric1} \bar{g}^{\mu\nu}_{+} & = g^{\mu\nu}-\dfrac{4 e^{\gamma}\sinh\gamma}{F}F^{\mu}_{\ \ \lambda}F^{\lambda\nu},\\
\label{eff_metric2} \bar{g}^{\mu\nu}_{-} & = g^{\mu\nu}.
\end{align}
These are the same metric tensors presented in Ref.~\cite{Guzman-Herrera:2023zsv} [see Eqs.~(3.4) and~(3.5) of the aforementioned paper, respectively]. Consequently, in spacetimes sourced by ModMax electrodynamics, one of the polarizations of light, namely $\mathcal{P}_{-}$, follows null geodesics according to the spacetime metric $g^{\mu\nu}$. Throughout this paper, we consider the effective metrics given by Eqs.~\eqref{eff_metric1} and~\eqref{eff_metric2}, unless otherwise stated. 

\section{BH solutions in ModMax theory}\label{sec:bhsmt}

In this section, we briefly review the derivation of static, spherically symmetric ModMax BHs with electric charge and some of their properties. The line element can be written as
\begin{align}
\nonumber ds^{2} & \equiv g_{\mu\nu}dx^{\mu}dx^{\nu}\\
\label{LE} & = -A(r)dt^{2}+B(r)dr^{2}+r^{2}d\Omega^{2},
\end{align}
where $A(r)$ and $B(r)$ are metric components to be determined by the field equations, and $d\Omega^{2} = d\theta^{2}+\sin^{2}\theta d\varphi^{2}$ is the line element of a unit 2-sphere. In NED, one can show that a spherically symmetric electromagnetic field satisfies~\cite{Dymnikova:2004zc}
\begin{equation}
\label{PCS}T_{0}^{\ 0} = T_{1}^{\ 1} \quad \text{and} \quad T_{2}^{\ 2} = T_{3}^{\ 3},
\end{equation}
as long as we consider a purely electrically (or magnetically) charged structure. Consequently, $G_{0}^{\ 0} = G_{1}^{\ 1}$ and we find that
\begin{equation}
A^{\prime}(r)B(r) = -A(r)B^{\prime}(r),
\end{equation}
where the prime denotes differentiation with respect to the radial coordinate $r$. Therefore, $A(r)B(r) = C$, where $C$ is a constant. One can show that $C = c^{2}$~\cite{hobson2006general}, but since we normalized $c = 1$, we get $A(r) = B(r)^{-1}$. Following the standard procedure to obtain BHs in NED, we also identify $A(r) = f(r)$, with $f(r)$ being a metric function defined as
\begin{equation}
\label{MF}f(r) \equiv 1 - \dfrac{2\mathcal{M}(r)}{r}.
\end{equation}
The function $\mathcal{M}(r)$ is determined by the ENED field equations, given by Eq.~\eqref{E-NED_F}, and its integration over the whole space provides the total mass $M$ of the central object~\cite{Fan:2016hvf}.

The only non-null components of the Faraday tensor are given by $F_{01}$ and $F_{10}$, which are related by 
\begin{equation}
F_{01} = -F_{10} = \dfrac{Q}{\mathcal{L}_{F}r^{2}},
\end{equation}
where $Q$ is a parameter related to the electric charge. Notice that the definition of the electric charge follows from the conservation of the Faraday tensor [cf. Eq.~\eqref{EFTC1}]. Thus, we get
\begin{equation}
\label{ecdef}Q_{e} = \dfrac{1}{4\pi}\int_{0}^{2\pi} \int_{0}^{\pi} \mathcal{L}_F F_{01} r^2 \sin\theta d\theta d\varphi.
\end{equation}
One can show that $Q = Q_{e}$. From the $G_{1}^{\ \ 1}$- and $G_{3}^{\ \ 3}$-components of the ENED field equations~\eqref{E-NED_F}, we find that
\begin{align}
\label{G11}\mathcal{M}^{\prime}(r) & = \dfrac{1}{4}\left[\mathcal{L}(F,G)-2F\mathcal{L}_{F}\right]r^{2}, \\
\label{G33}\mathcal{M}^{\prime\prime}(r) & = \dfrac{1}{2}\mathcal{L}(F,G)r, 
\end{align}
respectively. By solving Eq.~\eqref{G11} for $\mathcal{M}(r)$, we get
\begin{align}
\label{Q}\mathcal{M}(r) = C - \dfrac{e^{-\gamma}Q^{2}}{2r},
\end{align}
where $C$ is an integration constant. Since
\begin{equation}
\lim_{r \rightarrow \infty} \mathcal{M}(r) = M,
\end{equation}
we get $C = M$. Hence, the ModMax metric function leads to
\begin{equation}
\label{MF_EH}f(r) = 1 - \dfrac{2M}{r} + \dfrac{e^{-\gamma}Q^{2}}{r^{2}}.
\end{equation}

\begin{figure}[htbp]
\begin{centering}
    \includegraphics[width=\columnwidth]{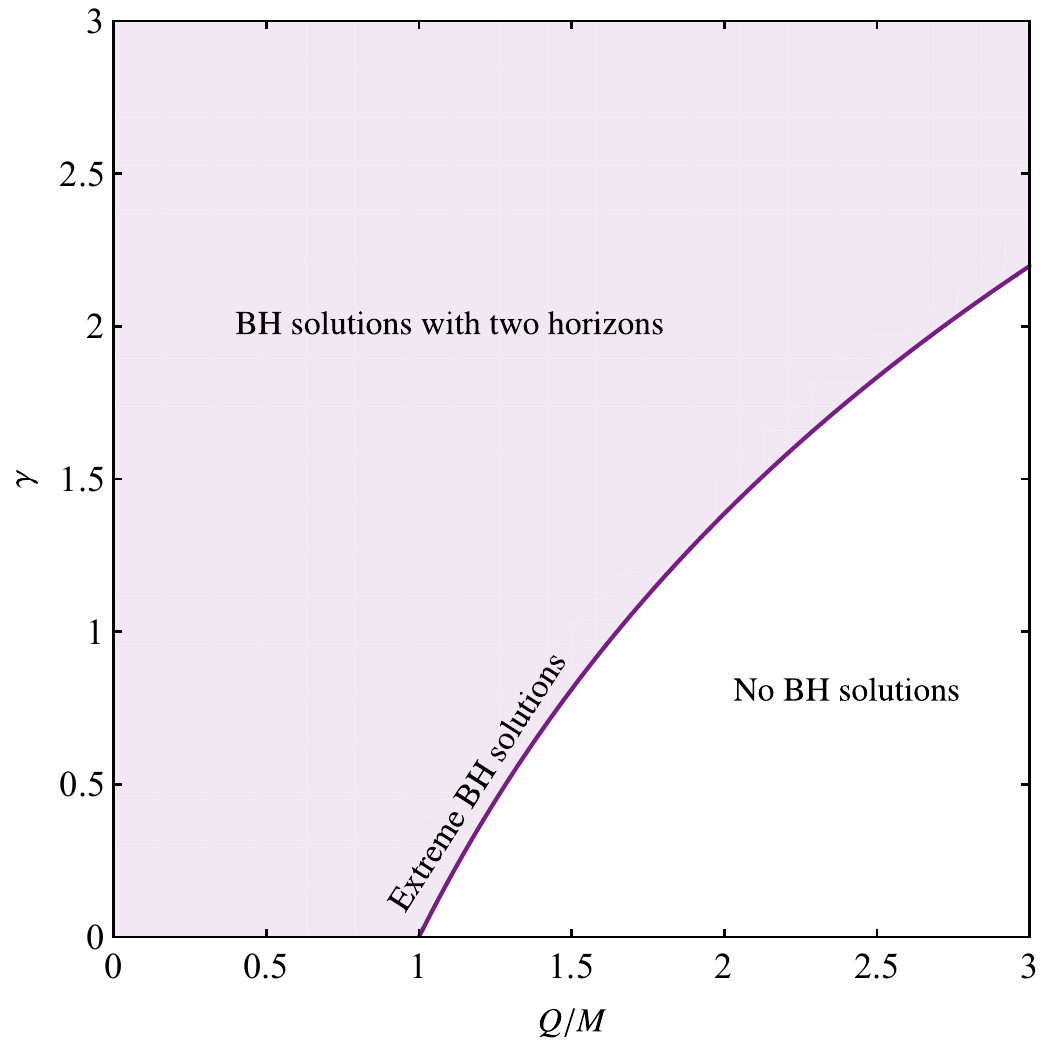}
    \caption{Horizon structure of the ModMax BH geometry as a $\gamma$ versus $Q/M$ two-dimensional map. In the map, we exhibit BH solutions with two horizons (purple shaded region), extreme BH solutions (solid purple curve), and no BH solutions (white region).}
    \label{existencelines}
\end{centering}
\end{figure}

The horizons can be found from $f(r) = 0$, namely,
\begin{equation}
\label{horizons}r_{\pm} = M \pm \sqrt{M^{2}-e^{-\gamma}Q^{2}},
\end{equation}
where $r_{+}$ and $r_{-}$ denote the event and Cauchy horizons, respectively. The extremal charge and the corresponding event horizon of an extremal charged ModMax BH are obtained by solving $f(r) = 0$ and $f^{\prime}(r) = 0$ simultaneously, leading to
\begin{equation}
\label{extcase}Q_{\rm{ext}} = e^{\gamma/2}M \quad \text{and} \quad r_{\rm{ext}} = M,
\end{equation}
respectively. In Fig.~\ref{existencelines}, we show the existence lines for BH solutions in Einstein-ModMax theory. We note that BH solutions are feasible when $0 \leq e^{-\gamma}Q^{2} \leq M^{2}$. Moreover, we obtain the RN spacetime when $\gamma = 0$, for which $Q_{\rm{ext}} = r_{\rm{ext}} = M$, and the Schwarzschild geometry for $Q = 0$.

\begin{figure}[htbp]
\begin{centering}
    \includegraphics[width=\columnwidth]{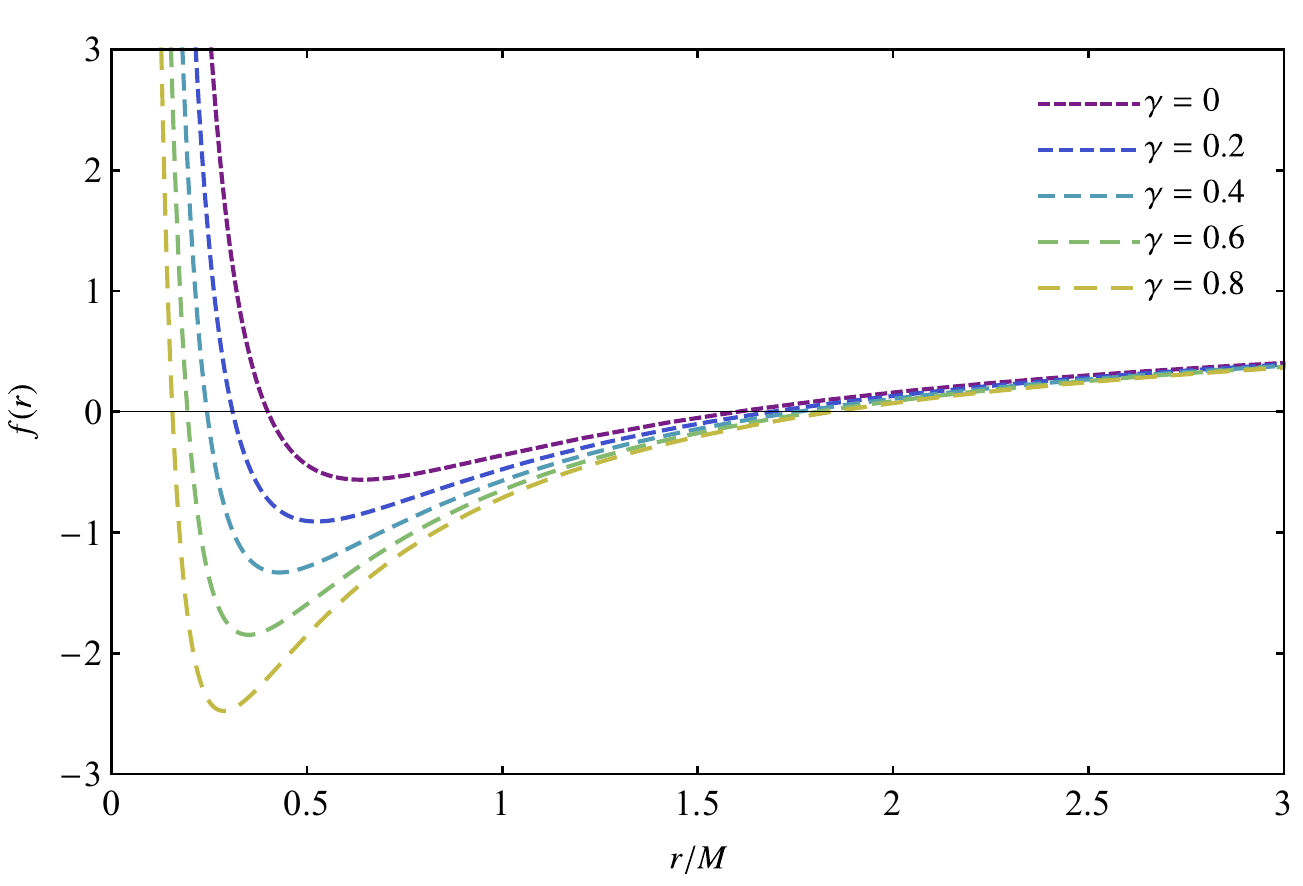}
    \caption{Metric function of the ModMax BH geometry, considering distinct values of $\gamma$, as a function of $r/M$. Here, we set $Q = 0.8M$.}
    \label{mfdifgamma}
\end{centering}
\end{figure}

In Fig.~\ref{mfdifgamma}, we display the behavior of the ModMax metric function. We observe that the size of the BH increases as we consider higher values of $\gamma$, for fixed values of $Q/M$. Therefore, Figs.~\ref{existencelines} and~\ref{mfdifgamma} show that the overall effect of $\gamma$ is to increase the BH size and the corresponding extremal charge value. We also point out that the ModMax BH geometry is singular, having a curvature singularity at $r = 0$~\cite{Guzman-Herrera:2023zsv}. Throughout this work, to better illustrate our main results, we typically consider values of $\gamma$ in the range $\gamma \in \left[0, 2\right]$. 

We end this section by presenting the effective geometry for photons in ModMax electrodynamics. The effective metrics can be obtained from Eqs.~\eqref{eff_metric1} and~\eqref{eff_metric2}. Since we are considering a purely electrically charged ModMax source, the line element for the first polarization of light results in
\begin{align}
\nonumber d\bar{s}^{2}_{+} & \equiv (\bar{g}_{\mu\nu})_{+}dx^{\mu}dx^{\nu}, \\
\label{LE_EF}& = e^{-2\gamma}\left[-f(r)dt^{2}+f(r)^{-1}dr^{2}\right]+r^{2}d\Omega^{2},
\end{align}
while for the second polarization of light, we find that
\begin{align}
\nonumber d\bar{s}^{2}_{-} & \equiv (\bar{g}_{\mu\nu})_{-}dx^{\mu}dx^{\nu}, \\
\label{LE_EF0}& =  -f(r)dt^{2}+f(r)^{-1}dr^{2}+r^{2}d\Omega^{2},
\end{align}
which is conformally related to the spacetime metric $g_{\mu\nu}$.

\section{Shapiro Time}\label{Shap}

In the following, we investigate the Shapiro time delay effect within the framework of ModMax electrodynamics. We first present the general formalism. Then we study the time delay effect. For completeness, we also present a numerical analysis of the total travel time and Shapiro delay.

\subsection{General Formalism}\label{subsec:gf}

For the static and spherically symmetric line element given by Eq.~\eqref{LE}, the trajectory of the light ray on the equatorial plane, $\theta=\pi/2$, can be written as~\cite{bhadra2010gravitational}
\begin{eqnarray}
    -A(r)\dot{t}^2+B(r)\dot{r}^2+r^2\dot{\varphi}^2=0\,.  \label{AB}
\end{eqnarray}
Notice that for the effective metrics [cf. Eqs.~\eqref{eff_metric1} and~\eqref{eff_metric2}], the angular sector is not modified by the NED fields. Thus, we keep $A(r)$ and $B(r)$ as general metric coefficients, following our Eq.~\eqref{LE}. The angular momentum is then given by
\begin{eqnarray}
    L=r^2\dot{\varphi}, \label{moment}
\end{eqnarray}
while the total energy reads 
\begin{eqnarray}
    E=A(r)\dot{t}\,.  \label{energy}
\end{eqnarray}
Using Eqs.\,\eqref{AB}-\eqref{energy}, the geodesic equation can be written as
\begin{eqnarray}
    \frac{d\varphi}{dr}=\frac{1}{r^2}\left[ \frac{1}{A(r)B(r)}\left(\frac{1}{b^2}-\frac{A(r)}{r^2}\right) \right]^{-1/2} \,. \label{geq}
\end{eqnarray}
where $b \equiv L / E$ is the impact parameter. Substituting Eqs.\,\eqref{moment}-\eqref{geq} into the geodesic equation \eqref{AB}, we get
\begin{equation}
    \frac{dt}{dr} = \frac{1}{b} \sqrt{\frac{B(r)}{A(r)}} \Biggl[ \frac{1}{b^2} - \frac{A(r)}{r^2} \Biggr]^{-1/2}\,,
    \label{eq:dtdr}
\end{equation}
where
\begin{equation}
    b = \sqrt{\frac{d^2}{A(d)}}\,
    \label{eq:bemd}
\end{equation}
is determined at the distance of closest approach of the light ray to the BH, $r=d$, where $dr/d\varphi = 0$. 
Accordingly, the total propagation time from $d$ to a point $r_i$ is given by
\begin{equation}
    t(r_i,d) = \frac{1}{b} \int_{d}^{r_i}   \sqrt{\frac{B(r)}{A(r)}} \Biggl[ \frac{1}{b^2} - \frac{A(r)}{r^2} \Biggr]^{-1/2}dr\,.
    \label{eq:tintegrado}
\end{equation}

We consider a light signal propagating between two observers located at radial positions $r_1$ and $r_2$. For a superior conjunction configuration, with $r_1 > r_2 \gg d$, where $d$ is the radius of closest approach of the photon trajectory to the BH, the total round-trip travel time can be written as
\begin{equation}
    T=2\left[t(r_1,d)+t(r_2,d)\right]. \label{totalt}
\end{equation}

The general formalism developed above enables the computation of the total propagation time of photons as well as the corresponding Shapiro time delay. The latter is defined by
\begin{equation}
    \Delta T\equiv T- T_0 \,,
\end{equation}
where $T_0$ is the corresponding travel time in flat spacetime. 
This formalism will be applied to the effective metrics \eqref{LE_EF} and \eqref{LE_EF0} associated with ModMax electrodynamics.  The resulting expressions for $T$ and $\Delta T$ for the two effective metrics will then be compared with those obtained for the RN spacetime, thereby quantifying the impact of NED fields on gravitational time-delay observables.

\subsection{Time delay effect in ModMax}\label{subsec:tdem}

As discussed in the previous section, two distinct effective geometries arise in ModMax electrodynamics, each governing a different polarization mode of light propagation. Therefore, in order to achieve a complete characterization of the time delay effect in  theory, we analyze the Shapiro delay separately for both effective metrics, comparing the behavior of light propagation in each of these sectors.

We first consider the propagation of photons. For radial null trajectories, the geodesic equation takes the form
\begin{eqnarray}
\label{geoeq}(\bar{g}_{\mu\nu})_{\pm} \left(\dot{x}^\mu\right)_{\pm} \left(\dot{x}^\nu\right)_{\pm} = 0\,,
\end{eqnarray}
where the four-vectors $\dot{x}^{\mu}$ are the contravariant components of the four-velocity. Notice that we need to index them with $\pm$, since vectors whose norm is null with respect to one polarization of light will not necessarily be null with respect to the other. Taking first the line element \eqref{LE_EF}, in \eqref{eq:tintegrado} we set $A(r)=e^{-2\gamma}f(r)$ and $B(r)=e^{-2\gamma}f(r)^{-1}$, obtaining 
\begin{eqnarray}
t_+(r,d)&=& \sqrt{r^2-d^2}+\frac{M\, \sqrt{r-d}}{\sqrt{r+d}}\nonumber\\
&&+4M\,\mathrm{arctanh}\left(\frac{\sqrt{r-d}}{\sqrt{r+d}}\right)\nonumber\\
&&-\frac{3Q^2\,e^{-\gamma}}{2d}\mathrm{arccot}\left(\frac{d}{\sqrt{r^2-d^2}}\right),
\end{eqnarray}
and therefore the total travel time \eqref{totalt} becomes
\begin{align}
T_+= \ & 2\Bigg[y_{1}+y_{2}+M\left(s_{1}+s_{2}\right)\nonumber\\
&+4M\left(\operatorname{arctanh}s_{1}+\operatorname{arctanh}s_{2}\right)\nonumber\\
&-\frac{3e^{-\gamma}Q^2}{2d}\left(\operatorname{arccot}\frac{d}{y_{1}}
+ \operatorname{arccot}\frac{d}{y_{2}}\right)\Bigg]. \label{T+}
\end{align}
where we defined
\begin{equation}
s_{i} \equiv \sqrt{\dfrac{r_{i}-d}{r_{i}+d}} \quad \text{and} \quad y_{i}=\sqrt{r_{i}^{2}-d^{2}}, 
\end{equation}
for simplicity, with $i = (1,2)$. Considering that $r_1>r_2\gg d$, the Shapiro time is 
\begin{equation}
\begin{split}
\Delta T_{+}
\simeq\,
&4M
\left[
1+
\ln\left(
\frac{4r_1 r_2}{d^2}
\right)
\right]
-\frac{3\pi e^{-\gamma}Q^2}{d}
\\
&\qquad
+3 e^{-\gamma}Q^2
\left(
\frac{1}{r_1}
+\frac{1}{r_2}
\right)
+\mathcal{O}\left(
\frac{d}{r_1},
\frac{d}{r_2}
\right).
\end{split}
\end{equation}

The obtained expression correctly reproduces the known limits of the theory. In the Maxwell limit, $\gamma\rightarrow 0$, the Shapiro delay reduces to the RN result \cite{Junior:2023nku}. Furthermore, the standard Schwarzschild result is recovered in the limit  $Q\rightarrow 0$. Therefore, the parameter $\gamma$ encodes the deviations induced by ModMax electrodynamics through exponential corrections to both the charge and mass contributions to the time delay.

We now consider the second effective metric \eqref{LE_EF0}. This effective geometry coincides with the standard ModMax spacetime. Proceeding as before, by substituting the corresponding metric functions into Eqs.\,\eqref{eq:tintegrado} and \eqref{totalt}, one finds that the total photon propagation time is identical to that obtained for the first effective metric. In other words, we find that
\begin{eqnarray}
   \Delta T_{\rm ModMax}= \Delta T_{+}=\Delta T_{-}\,.
\end{eqnarray}

This result deserves particular attention. Although the two effective metrics correspond to distinct polarization modes of the electromagnetic field in ModMax electrodynamics, they predict exactly the same total propagation time and, consequently, the same Shapiro time delay. At first sight, this equality is rather unexpected, since the effective geometries differ in their temporal and radial sectors. However, as shown explicitly in Appendix~\ref{appendix_invariance}, the invariance of the propagation time follows from the specific structure of the null-geodesic equation governing the coordinate time. In particular, the simultaneous constant rescaling of the metric functions entering the temporal and radial sectors leaves the integrand of Eq.~\eqref{eq:tintegrado} unchanged after the corresponding rescaling of the impact parameter. As a consequence, the propagation time is identical for both polarization modes. This demonstrates that, within the geometric optics approximation, the Shapiro time delay is insensitive to the birefringent structure of ModMax electrodynamics, despite the existence of two distinct effective metrics. Therefore, no polarization dependent time delay is expected in this configuration, implying that the Shapiro effect alone cannot be used to discriminate between the two effective geometries. This indicates that any observational signature of birefringence according to the ModMax BH must be sought in other null geodesic observables, e.g., the Sagnac effect (cf. Sec.~\ref{sec:se}) and the kinematic shifts (cf. Sec.~\ref{sec:gkr}), which are sensitive to the differences between the effective metrics.

\subsection{Numerical analysis of the total travel time and Shapiro delay}\label{subsec:natt}

In what follows, we compare the predictions of the effective ModMax geometry with those of the RN spacetime.

For the stellar mass BH \cite{Bambi:2025rod}, we adopt $M=6.1M_\odot$, $Q=0.5M$, $d=100M$, $r_1=2\times10^4d$, and $r_2=10^4d$. The corresponding Shapiro time delays for different values of the ModMax parameter are listed in Table~\ref{tab:stellar}. The procedure used to convert the Shapiro time delays from geometrized units to seconds is detailed in Appendix\, \ref{app:units}. As expected, the RN result is recovered for $\gamma=0$, while increasing $\gamma$ produces only a slight increase in the delay, reaching a maximum relative deviation of $2.60\times10^{-4}$ for $\gamma=3$.
\begin{table}[!ht]
\centering
\begin{tabular}{|c|c|c|c|}
\hline
$\gamma$ & $\Delta T_{\rm ModMax}$ (s) & $\Delta T_{\rm RN}$ (s) & Relative difference \\
\hline
0 & 0.00258322 & 0.00258322 & 0 \\
\hline
0.5 & 0.0025835 & 0.00258322 & 0.000107825 \\
\hline
1 & 0.00258367 & 0.00258322 & 0.000173224\\
\hline
1.5 & 0.00258377 & 0.00258322 & 0.000212891 \\
\hline
2 & 0.00258383 & 0.00258322 & 0.00023695 \\
\hline
2.5 & 0.00258387 & 0.00258322 & 0.000251542 \\
\hline
3 & 0.00258389 & 0.00258322 & 0.000260393 \\
\hline
\end{tabular}
\caption{Shapiro time delay predicted by the effective ModMax geometry and the corresponding RN spacetime for a stellar mass BH with $M=6.1M_\odot$ and $Q=0.5M$.}
\label{tab:stellar}
\end{table}

The same analysis is performed for the supermassive BH Sgr A$^\star$ \cite{ghez2008measuring,boehle2016improved,event2022first}, adopting $M=4.25\times10^6M_\odot$ and the same dimensionless parameters. The numerical results are presented in Table~\ref{tab:sgrA}. Although the absolute Shapiro delay is several orders of magnitude larger due to the larger BH mass, the relative deviations from the RN solution are identical to those of the stellar mass case, reflecting the fact that all characteristic distances scale linearly with $M$.
\begin{table}[!ht]
\centering
\begin{tabular}{|c|c|c|c|}
\hline
$\gamma$ & $\Delta T_{\rm ModMax}$ (s) & $\Delta T_{\rm RN}$ (s) & Relative difference \\
\hline
0 & 1799.78 & 1799.78 & 0 \\
\hline
0.5 & 1799.98 & 1799.78 & 0.000107825 \\
\hline
1 & 1800.1 & 1799.78 & 0.000173224  \\
\hline
1.5 & 1800.17 & 1799.78 & 0.000212891 \\
\hline
2 & 1800.21 & 1799.78 & 0.00023695\\
\hline
 2.5 & 1800.24 & 1799.78 & 0.000251542\\
\hline
3 & 1800.25 & 1799.78 & 0.000260393\\
\hline
\end{tabular}
\caption{Shapiro time delay predicted by the effective ModMax geometry and the corresponding RN spacetime for Sgr A$^{\star}$, with $M=4.25\times10^{6}M_\odot$ and $Q=0.5M$.}
\label{tab:sgrA}
\end{table}

These results show that the ModMax corrections remain extremely small throughout the weak field regime. As follows from Eq.~\eqref{T+}, the deviations from the RN prediction are entirely governed by the exponential factor $e^{-\gamma}$. As $\gamma$ increases, the charge dependent contribution is progressively suppressed, leading to a slight increase in the total Shapiro time delay with respect to the RN spacetime. Consequently, even for relatively large values of $\gamma$, the relative differences remain at the level of $10^{-4}$, indicating that the Shapiro time delay alone is only weakly sensitive to nonlinear electromagnetic effects in the physical scenarios considered here.

\section{Sagnac effect}\label{sec:se}

In this section, we investigate the Sagnac effect in detail. We first review the Sagnac time. Then we provide a general framework to describe this effect in NED-sourced spacetimes, considering the effective metrics. As a proof of concept, we discuss the application of this framework to the ModMax BH. 

\subsection{Preliminaries}\label{subsec:preli}

\begin{figure}[!htbp]
\begin{centering}
    \includegraphics[width=\columnwidth]{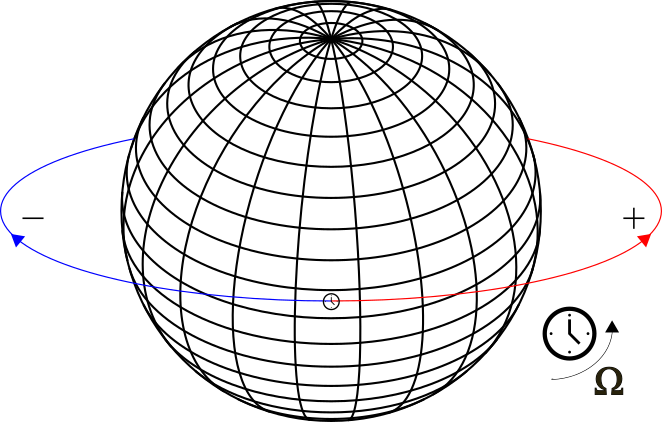}
    \caption{Illustration of the setup associated with the Sagnac effect. We consider the proper time difference between the emission and absorption of the co-propagating ($+$) and counter-propagating ($-$) light beams, considering the effective metrics. The proper time is measured by a clock rotating with constant angular velocity $\Omega$ with respect to the BH. This figure is based on Fig. 1 of Ref.~\cite{Ruggiero:2005nd}. }
    \label{sagnaceffect}
\end{centering}
\end{figure}

The Sagnac effect consists of a source emitting a light beam that splits into two beams traveling in opposite directions along a rotating circular platform. These beams eventually return to the source at different times. For simplicity, the two beams have the same velocity (in absolute value) with respect to the rotating frame, and the beams are spinless. In Fig.~\ref{sagnaceffect}, we present a schematic illustration of the Sagnac effect.

The proper time difference is measured by a source, which we identify as a clock rotating with a constant angular velocity $\Omega$ with respect to the standard geometry [cf. Eq.~\eqref{LE}], assuming that the observer (the source or clock) is a neutral, massive object. Moreover, the Sagnac effect in curved spacetimes can be obtained through a formal analogy with the Aharonov-Bohm effect~\cite{Ashtekar:1975wt,Ruggiero:2005nd} or by investigating the corresponding line element of the central object~\cite{Souza:2024ltj}. These two methods are equivalent (see, e.g., Ref.~\cite{rizzi2004relativistic}), but here we use the latter to benefit from the general line element used in Sec.~\ref{Shap}. 

To obtain the Sagnac effect, we first rewrite the general line element given by Eq.~\eqref{LE}, considering a transformation to a rotating frame with angular velocity $\Omega$ and angular displacement given by $\varphi = \varphi_{0}-\Omega t$. Thus, we perform the transformation  $\varphi \rightarrow \varphi_{0}-\Omega t$ in Eq.~\eqref{LE}, suppressing the subscript $0$ in $\varphi_{0}$ for simplicity. We also assume that the rotation occurs in the equatorial plane at a constant radius $R$. Thus, we get
\begin{equation}
\label{LE_general_sagnac} ds^{2} = -\left[A(R)-\dfrac{\Omega^2 R^{2}}{c^{2}} \right]c^{2}dt^{2}-2\Omega R^{2}d\varphi dt + R^{2}d\varphi^{2}.
\end{equation}
We also retrieve the constant $c$. Notice that Eq.~\eqref{LE_general_sagnac} is equivalent to Eqs.~(6) and~(72) of Refs.~\cite{Souza:2024ltj,Fathi:2019jgd}, respectively.

The trajectory of light is obtained by setting $ds^{2} = 0$ in Eq.~\eqref{LE_general_sagnac}. Moreover, notice that for the ModMax BH, the contributions of the effective metrics are encoded in $A(R)$ [cf. Eqs.~\eqref{LE_EF} and~\eqref{LE_EF0}]. Since our goal is to obtain the time difference between the co- and counter-propagating light beams, we solve $ds^{2} = 0$ for $dt$, leading to
\begin{equation}
\label{dtgeneral}dt = \dfrac{\left(-\Omega R \pm c\sqrt{A(R)} \right)}{\Omega^{2}R^{2}-c^{2}A(R)}R d \varphi.    
\end{equation}
To properly identify the light paths followed by each light beam, we need to analyze the denominator of Eq.~\eqref{dtgeneral}. The timelike condition imposed on the observer implies that $c^{2}A(r)>\Omega^{2}R^{2}$. In other words, the angular velocity of the source is always smaller than the effective angular velocity of the light beams. Therefore, for co-propagating photons, for which $d\varphi > 0$, we take the negative root of Eq.~\eqref{dtgeneral}. Conversely, for counter-propagating photons, for which $d\varphi < 0$, we take the positive root. These conditions ensure  that the photons are traveling forward in time. Thus, we find that
\begin{align}
\label{tco}dt_{+} & = \dfrac{R}{c\sqrt{A(R)}-\Omega R}d\varphi, \\ 
\label{tcounter}dt_{-} & = \dfrac{R}{c\sqrt{A(R)}+\Omega R}d\varphi,
\end{align}
for the co- and counter-propagating light beams, respectively. The time interval difference between the light beams is defined by $\Delta t \equiv t_{+}-t_{-}$, leading to
\begin{equation}
\Delta t = 4\pi \dfrac{\Omega R^{2}}{c^{2}A(R)-\Omega^{2}R^{2}}.
\end{equation}

As the Sagnac effect is the measurement of the time delay using a physical detector, we need to consider the time difference as measured by a source, which is a timelike observer measuring the invariant proper time $\Delta \tau$ defined by
\begin{equation}
\Delta \tau \equiv \sqrt{A(R)-\dfrac{\Omega^2 R^{2}}{c^{2}} }\Delta t.
\end{equation}
Therefore, the Sagnac time can be written as
\begin{equation}
\label{sagnactime} \Delta \tau = \dfrac{4\pi}{\Omega_{R}}\dfrac{\Omega}{\sqrt{\Omega_{0}^{2}-\Omega^{2}}},
\end{equation}
where
\begin{equation}
\label{freqsignac}\Omega_{R} \equiv \dfrac{c}{R} \quad \text{and} \quad \Omega_{0} \equiv \dfrac{c\sqrt{A(R)}}{R}
\end{equation}
are defined as the angular velocity of light in flat spacetimes and the effective angular velocity of light in curved spacetimes, respectively. Notice that Eq.~\eqref{sagnactime} reduces to Eq. (78) of Ref.~\cite{Fathi:2019jgd}, provided that we identify $A(r)$ as the $g_{tt}$ component of the metric tensor of the charged Weyl BH. Moreover, by presenting the Sagnac effect in the form given by Eq.~\eqref{freqsignac}, we isolate gravitational effects entirely inside $\Omega_{0}$. We also point out that the causality constraint reduces to $\Omega < \Omega_{0}$; otherwise, the Sagnac time leads to nonphysical values.

\subsection{Sagnac effect in nonlinear electrodynamics}\label{subsec:sened}

As discussed earlier, photons in NED follow null geodesics according to the effective metrics given by Eqs.~\eqref{LE_EF} and~\eqref{LE_EF0}. If we consider counter-propagating light beams on the equatorial plane ($\theta = \pi/2$) along fixed circular trajectories with a radius $r = R$, one can show that $A(R)$ can be written as
\begin{align}
A(R)_{+} & \equiv  e^{-2\gamma} f(R), \\
A(R)_{-} &\equiv f(R),
\end{align}
for the light polarizations $\mathcal{P}_+$ and $ \mathcal{P}_-$, respectively. Therefore, the Sagnac effect for each polarization of light is given by Eq.~\eqref{sagnactime}, with the effective angular velocities satisfying
\begin{align}
\label{angvecplus} \Omega_{0}^{+} & = e^{-\gamma}\sqrt{\Omega_{R}^{2}-\Omega_{M}^{2}+\Omega_{Q}^{2}} = e^{-\gamma} \Omega_0^-,\\
\label{angvecminus} \Omega_{0}^{-} & = \sqrt{\Omega_{R}^{2}-\Omega_{M}^{2}+\Omega_{Q}^{2}},
\end{align}
for the light polarizations $\mathcal{P}_+$ and $ \mathcal{P}_-$, respectively, where
\begin{equation}
\label{angvecs}\Omega_{M} \equiv \dfrac{c}{R}\sqrt{\dfrac{2M}{R}} \quad \text{and} \quad \Omega_{Q} \equiv \dfrac{e^{-\gamma/2} Q c}{R^{2}}.
\end{equation}
Notice that the results for the polarization $\mathcal{P}_-$ coincide with the results obtained in standard geometry, as this light polarization is conformally related to the spacetime metric. Eqs.~\eqref{angvecplus} and~\eqref{angvecminus} can be understood as follows. The angular velocities $\Omega_{M}$ and $\Omega_{Q}$ are contributions to the angular velocity of light $\Omega_{0}$ in curved spacetimes associated with the mass and charge of the charged BH, respectively. In particular, we observe that for the light polarization $\mathcal{P}_+$ there is a scalar factor given by $e^{-\gamma}$ shifting the angular velocity of light obtained according to the light polarization $\mathcal{P}_-$. Moreover, for $\Omega = 0$, i.e., in a non-rotating frame, we find that $dt_{+} = dt_{-}$, implying that the propagation is symmetrical in both directions. In other words, there is no Sagnac effect ($\Delta \tau = 0$).

\subsection{Sagnac effect in ModMax geometry}\label{subsec:semg}

In what follows, we focus on using the Sagnac effect to explore NED-sourced BHs, considering, in particular, the ModMax BH geometry from three complementary perspectives. First, to establish lower bounds on the NED parameter $\gamma$. Second, to examine how much the NED results deviate from the linear electrodynamics results (i.e., the results for the RN BH spacetime). Third, to investigate the influence of the BH parameters, in particular $Q$ and $\gamma$, on the Sagnac effect.

\begin{figure}[!htbp]
\begin{centering}
    \includegraphics[width=\columnwidth]{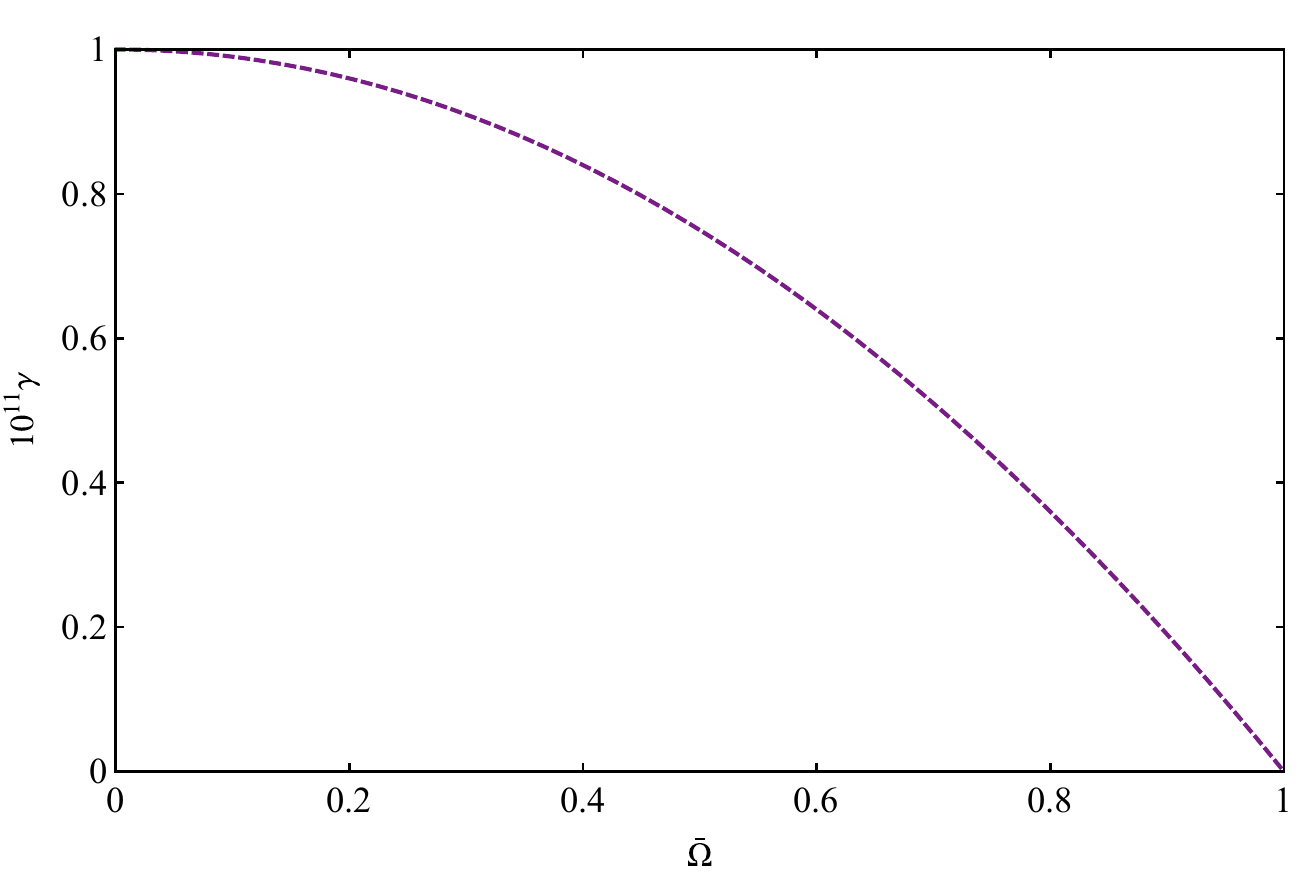}
    \caption{Allowed values for $\gamma$, as a function of $\bar{\Omega}$, considering the restriction imposed by Eq.~\eqref{constgamma} on the ModMax parameter $\gamma$.}
    \label{constraintgamma}
\end{centering}
\end{figure}

The relative deviation between the Sagnac effect for each polarization of light is defined by
\begin{equation}
\label{sagnacdev}\Delta_{\gamma} \equiv \dfrac{\left|\Delta\tau_+-\Delta\tau_-\right|}{\Delta\tau_-} = \left|\dfrac{\sqrt{1-\bar{\Omega}^{2}}}{\sqrt{e^{-2\gamma}-\bar{\Omega}^{2}}}-1\right|,
\end{equation}
where $\bar{\Omega} \equiv \Omega/\Omega_{0}^-$. Notice that this expression holds only if $\Omega < \Omega_0^{-}$, which is to be expected as the Sagnac effect must be a real number, and $e^{-2\gamma} > \bar{\Omega}^{2}$ to avoid nonphysical results. Moreover, we assume that the parameters $M$, $Q$, $\gamma$, and $R$ are the same for both light polarizations.  In Ref.~\cite{Souza:2024ltj}, the authors argue that if the relative Sagnac time deviation is outside the limit of clock accuracy, defined by  $\Delta_{c} \equiv 10^{-11}$~\cite{jaduszliwer2021past}, then constraining the influence of the gravitational system through the Sagnac effect is feasible. Thus, we impose that $\Delta_\gamma \geq \Delta_c$. By combining this condition with Eq.~\eqref{sagnacdev}, we get
\begin{equation}
\label{constgamma}\gamma \geq - \dfrac{1}{2}\ln \left[\dfrac{1+\bar{\Omega}^{2}\Delta_c(2+\Delta_c)}{(1+\Delta_c)^{2}} \right].
\end{equation}
This is the only physical solution for the algebraic system formed by Eq.~\eqref{sagnacdev} and $\Delta_\gamma \geq \Delta_c$, as $\gamma \geq 0$ to avoid causality issues. In Fig.~\ref{constraintgamma}, we display Eq.~\eqref{constgamma} as a function of $\bar{\Omega}$. We observe that the minimum required value of $\gamma$ for the Sagnac effect deviation to be detectable is very small and diminishes as we increase $\bar{\Omega}$, with the lower bound of $\gamma \approx 10^{-11}$ for $\bar{\Omega} \rightarrow 0$. In other words, for ModMax BHs with $\gamma > 10^{-11}$, it would be possible to measure the relative deviation between the Sagnac effect for each light polarization, considering the same BH parameters.

From an observational perspective, this result suggests that we can use the Sagnac effect as a mechanism for measuring vacuum birefringence in gravitational setups. The fundamental requirement is an astrophysical object endowed with an electromagnetic field sufficiently strong to trigger NED effects, along with an experimental apparatus capable of measuring the Sagnac effect. This could be potentially relevant for some magnetars due to their intense magnetic fields~\cite{Kennea:2013dfa,Olausen:2013bpa}.

We now focus on the relative deviation between the Sagnac effect for each light polarization with respect to the RN BH. For simplicity, we consider that the parameters $M$, $Q$, $\gamma$, and $R$ are the same for $\Delta\tau_{\pm}$ and $\Delta\tau_{\rm{RN}}$. Thus, we find that
\begin{equation}
\label{sagnacdev2}\Delta^{\pm} \equiv \dfrac{\left|\Delta\tau_{\pm}-\Delta\tau_{\rm{RN}}\right|}{\Delta\tau_{\rm{RN}}} = \left|\dfrac{\sqrt{1-\bar{\Omega}^{2}_{0}}}{\sqrt{\left(\Omega_{0}^{\pm}/\Omega_{0}\right)^2-\bar{\Omega}^{2}_{0}}}-1\right|,
\end{equation}
where $\bar{\Omega}_{0} \equiv \Omega/\Omega_{0}$, with $\Omega_0$ being the effective angular velocity for the RN BH . This expression holds provided that $\Omega < \Omega_{0}$, and for $\Delta^{+}$ we have the additional requirement that $e^{-2\gamma} > \bar{\Omega}^{2}_{0}$. Notice that the only relevant comparison comes from $\Delta^{+}$, since for $\Delta^{-}$, we observe that $\Delta\tau_- \approx \Delta \tau_{\rm{RN}}$. Although they differ analytically by the factor $e^{-\gamma}$ in the charge-to-mass sector, this contribution is subleading. Therefore, the results for the Sagnac effect in the $\mathcal{P}_{-}$-polarization are practically indistinguishable from those obtained in the RN case, considering that the parameters describing both BH geometries coincide, namely, the mass and charge of the BHs.

\begin{figure}[!htbp]
\begin{centering}
    \includegraphics[width=\columnwidth]{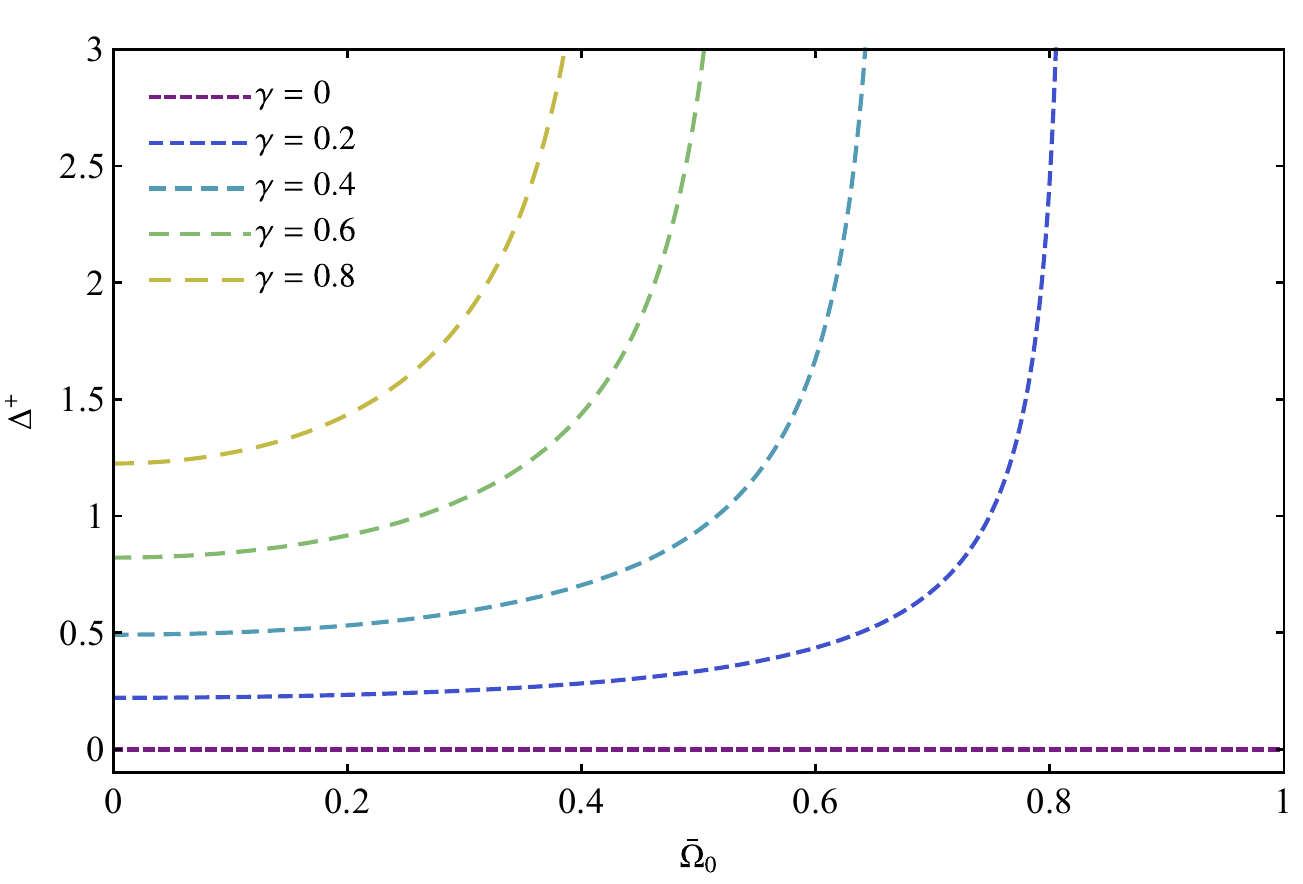}
    \includegraphics[width=\columnwidth]{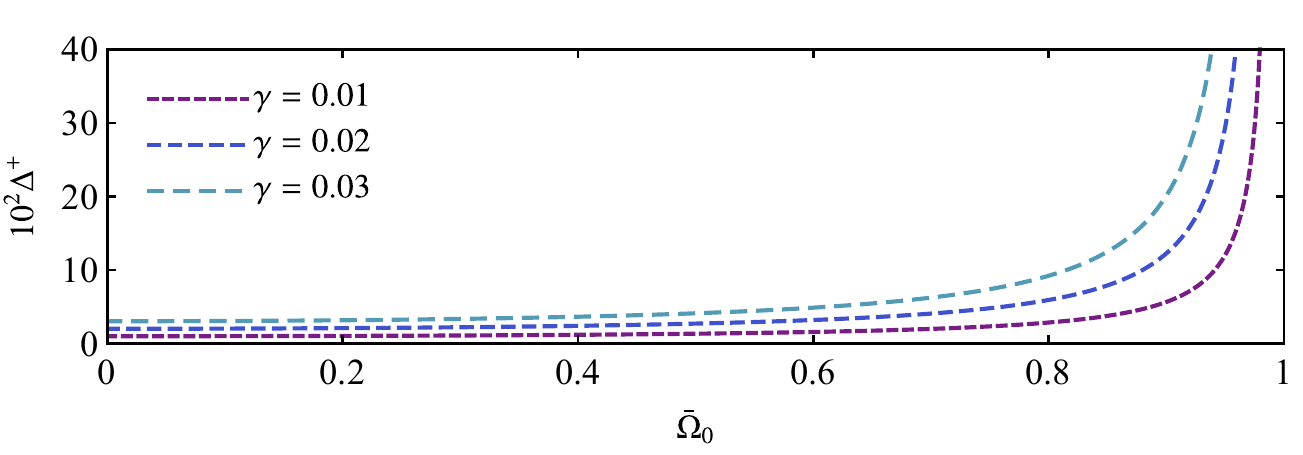} 
    \caption{Top panel: Relative deviation between the Sagnac effect for the light polarization $\mathcal{P}_{+}$ and that for the RN case, as a function of $\bar{\Omega}_{0}$, considering distinct values of $\gamma$. For each value of $\gamma$, $\bar{\Omega}_{0}$ must be smaller than $1$, $0.8187$, $0.6703$, $0.5488$, and $0.4493$, respectively. Bottom panel: Same as the top panel, but considering smaller values of $\gamma$. In this case, for each value of $\gamma$, $\bar{\Omega}_{0}$ must be smaller than $0.99005$, $0.98019$, and $0.97044$, respectively.}
    \label{Deltaplus}
\end{centering}
\end{figure}

In Fig.~\ref{Deltaplus}, we show $\Delta^{+}$ as a function of $\bar{\Omega}_{0}$. We see that for the light polarization $\mathcal{P}_{+}$, the relative deviation of the Sagnac effect with respect to the RN BH increases as we consider higher values of $\gamma$. We also point out that even a tiny value of $\gamma$ can lead to considerably large differences, provided that we have sufficiently fast sources. For example, for $\gamma = 0.01$ and $\bar{\Omega}_{0} = 0.95$, the difference is around $12\%$.

\begin{figure}[!htbp]
\begin{centering}
    \includegraphics[width=\columnwidth]{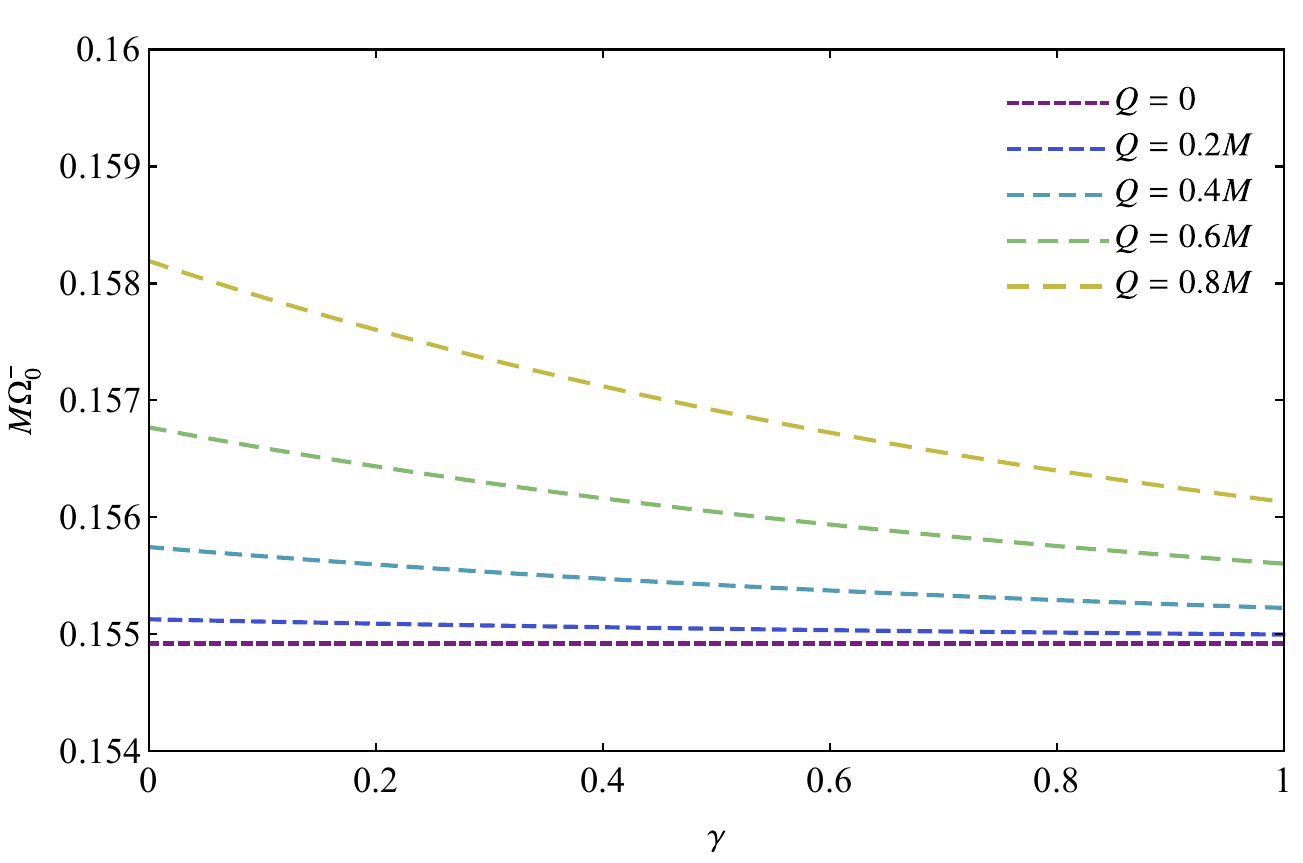}
    \caption{Effective angular velocity of photons according to the polarization $\mathcal{P}_{-}$, as a function of $\gamma$, considering distinct values of $Q/M$. Here, for simplicity, we set $R = 5M$ and $c = 1$. }
    \label{angvecminusfig}
\end{centering}
\end{figure}

To verify the influence of the charged BH parameters $Q$ and $\gamma$ on the Sagnac effect, we investigate the behavior of the angular velocity of photons in both light polarizations for a fixed $R$. If $\Omega_{0}^{\pm}$ increases (decreases), for fixed values of $\Omega_{R}$ and $\Omega$, then $\Delta \tau_{\pm}$ decreases (increases) [cf. Eq.~\eqref{sagnactime}]. Physically, this means that if the light has a higher (lower) effective angular velocity $\Omega_0$, it completes the circuit faster (slower), giving the observer less (more) time to move during the transit. This results in a smaller (larger) proper time difference. In Fig.~\ref{angvecminusfig}, we display the behavior of $\Omega_{0}^{-}$. As we can see, the effective angular velocity  $\Omega_{0}^{-}$ increases as we consider higher values for the charge-to-mass ratio but decreases as we increase $\gamma$. Notice that since $\Omega_{0}^{+} = e^{-\gamma} \Omega_0^-$ [cf. Eq.~\eqref{angvecplus}], the effect of $\gamma$ on the angular velocity of the photon for the first polarization $\mathcal{P}_{+}$ is to further reduce the angular velocity. Therefore, we see that the BH charge contributes to the decrease of the Sagnac effect, while the NED parameter contributes to its increase.

\section{Gravitational and kinematic redshifts}\label{sec:gkr}

In this section, we investigate the gravitational and kinematic redshifts in the background of the ModMax BH. We first introduce the main equations necessary to investigate the redshifts. Then, we particularize the equations for the ModMax case, considering the effective geometries.

\subsection{Framework}\label{subsec:framework}

We once again consider that the photon is governed by the null geodesic equation in the effective metric [cf. Eq.~\eqref{geoeq}]: 
\begin{eqnarray}
    \left(\bar{g}_{\mu\nu}\right)_{\pm}k^\mu_{\pm} k^\nu_{\pm}=0\,,\label{eqk}
\end{eqnarray}
where $k^\mu_{\pm}=(k^t_{\pm},k^r_{\pm}, k^\theta_{\pm}, k^\varphi_{\pm})$ is the 4-wave vector\footnote{To write the equations in a general form and in accordance with other works in the literature [see, e.g., Refs.~\cite{martinez2024observational,de2025exploring,Guzman-Herrera:2023zsv}], we express both the usual metric and the effective metric in terms of arbitrary metric coefficients, i.e., $ds^{2} = g_{tt}dt^{2}+g_{rr}dr^{2}+g_{\theta\theta}d\theta^{2}+g_{\varphi\varphi}d\varphi^{2}$ for the standard geometry. For the effective geometry, the same holds with the replacement $g \rightarrow \bar{g}$.\label{nota}}. Notice that we have introduced the indices $\pm$ to take into account both polarizations of light. For the blueshift/redshift problem, we consider a photon emitting source $e$ in a circular orbit around the BH, confined to the equatorial plane $\theta=\pi/2$, for which the $k^\theta_{\pm}$ component vanishes, and a receiver with radial coordinate $d$. The corresponding 4-velocities are given by
\begin{equation}
u_e^\mu=(u^t,0,0,u^\varphi)_e \quad \text{and} \quad u_d^\mu=(u^t,0,0,u^\varphi)_d,
\end{equation}
respectively. The indices $e$ and $d$ denote that the quantities are measured according to the emitter and detector, respectively, which follow timelike geodesics. Furthermore, the background spacetime possesses symmetries described by the Killing vectors $\xi^\mu=(1,0,0,0)$ and $\chi^\mu=(0,0,0,1)$, which allow us to define the conserved energy  and angular momentum associated with the photon motion, respectively, as
\begin{eqnarray}
\bar{E}_{\pm}=-\left(\bar{g}_{\mu\nu}\right)_{\pm}\xi^\mu k^\nu_{\pm} \quad \text{and}  \quad \bar{L}_{\pm}=\left(\bar{g}_{\mu\nu}\right)_{\pm}\chi^\mu k^\nu_{\pm} \,, \label{EL}
\end{eqnarray}

The frequencies of the emitted and absorbed photons are, respectively, given by
\begin{eqnarray}
    \left(\omega_e\right)_{\pm}=-g_{\mu\nu}k^\mu_{\pm} u^\nu_e \quad \text{and}  \quad \left(\omega_d\right)_{\pm}=-g_{\mu\nu}k^\mu_{\pm} u^\nu_d\,. \label{freq}
\end{eqnarray}
Now, using Eqs. \eqref{EL} and \eqref{freq}, together with the considerations discussed above, the expression for evaluating the gravitational and kinematic shift in static, spherically symmetric backgrounds is given by
\begin{eqnarray}
    1+z_{\pm}=\frac{\left(\omega_e\right)_{\pm}}{\left(\omega_d\right)_{\pm}}=\frac{\left[ \bar{\mathcal{A}}_{\pm}\,\bar{E}_{\pm} u^t-\bar{\mathcal{D}}_{\pm}\,\bar{L}_{\pm} u^\varphi\right]_e}{\left[\bar{\mathcal{A}}_{\pm}\,\bar{E}_{\pm} u^t-\bar{\mathcal{D}}_{\pm}\,\bar{L}_{\pm} u^\varphi\right]_d}\,,\label{redshift}
\end{eqnarray}
where $\bar{\mathcal{A}}_\pm=g_{tt}/\left(\bar{g}_{tt}\right)_\pm$ and $\bar{\mathcal{D}}_\pm=g_{\varphi\varphi}/\left(\bar{g}_{\varphi\varphi}\right)_\pm$.   
We now consider three distinct situations. First, we consider the case where both the emitting source and the receiver, located at an arbitrary distance, are static ($u^\varphi=0$). Therefore, the redshift is purely gravitational, and Eq. \eqref{redshift} reduces to
\begin{eqnarray}
    1+\left(z_{\rm grav}\right)_{\pm}=\frac{\left[\bar{\mathcal{A}}_{\pm}\left(-g_{tt}\right)^{-1/2}\right]_e}{\left[\bar{\mathcal{A}}_{\pm}\left(-g_{tt}\right)^{-1/2}\right]_d}\,,\label{zgrav}
\end{eqnarray}
where we have used the relation $u^t=\left(-g_{tt}\right)^{-1/2}$. The second case is a particular case of the first, in which we consider the receiver to be located very far away, yielding $u^t_d=1$ and $(g_{tt})_d \rightarrow -1$. Therefore, Eq. \eqref{redshift} takes the form
\begin{eqnarray}
   1+\left(z_{\rm grav}\right)_{\pm}= \frac{\left[\bar{\mathcal{A}}_{\pm}\left(-g_{tt}\right)^{-1/2}\right]_e }{\left[-(\bar{g}_{tt})_{\pm}\right]^{-1}_d}\,. \label{zgrav2}
\end{eqnarray}

As a third case, we consider a receiver (detector) located very far from a source orbiting the BH. Before proceeding, we introduce the photon's impact parameter \cite{Martinez-Valera:2023guj}, defined as $\bar{b}_{\pm}= \bar{L}_{\pm}/\bar{E}_{\pm}$, which can also be written as 
\begin{eqnarray}
    \bar{b}_{\pm}=\pm\sqrt{-\dfrac{\left(\bar{g}_{\varphi\varphi}\right)_{\pm}}{\left(\bar{g}_{tt}\right)_{\pm}}}\,.
\end{eqnarray}
Using this result together with the expressions for the 4-velocity given in \cite{Martinez-Valera:2023guj},
\begin{eqnarray}
u^t_e=-\frac{E}{g_{tt}}=\sqrt{\frac{g'_{\varphi\varphi}}{g'_{tt}g_{\varphi\varphi}-g'_{\varphi\varphi}g_{tt}}}
\end{eqnarray}
and 
\begin{eqnarray}
    u^\varphi_e=\frac{L}{g_{\varphi\varphi}}=\sqrt{\frac{g'_{tt}}{g_{tt}g'_{\varphi\varphi}-g_{\varphi\varphi}g'_{tt}}}\,,
\end{eqnarray}
where $E$ and $L$ are the energy and angular momentum of a star treated as a particle following a circular geodesic (see, e.g., Ref.~\cite{de2025exploring}). Consequently, one can show that Eq. \eqref{redshift} now yields the total blueshift ($+$) and redshift ($-$), namely,
\begin{widetext}
\begin{equation}
1+\left(z_{\rm tot}^{\mathcal{P}_{\pm}}\right)_{\pm}=\left[\mathcal{\bar{A}}_{\pm}\sqrt{\frac{g'_{\varphi\varphi}}{g'_{tt}g_{\varphi\varphi}-g'_{\varphi\varphi}g_{tt}}}\mp \mathcal{\bar{D}}_{\pm}\sqrt{-\dfrac{\left(\bar{g}_{\varphi\varphi}\right)_{\pm}}{\left(\bar{g}_{tt}\right)_{\pm}}}\sqrt{\frac{g'_{tt}}{g_{tt}g'_{\varphi\varphi}-g_{\varphi\varphi}g'_{tt}}}\right]_{e}\left[-\left(\bar{g}_{tt}\right)_{\pm} \right]_{d}. \label{redgeral1}
\end{equation}
\end{widetext}

We emphasize that to derive Eq.~\eqref{redgeral1}, we have considered that the general static, spherically symmetric spacetime is asymptotically flat. Otherwise, we could not assume that $\left(g_{tt}\right)_{d} \rightarrow -1$ far away from the BH. Moreover, Eq.~\eqref{redgeral1} provides the general formulas for the total blueshift and redshift, taking into account the effective metrics. We notice that in many works, the notation $\pm$ is used to denote the blueshift ($+$) and redshift ($-$). Nevertheless, in our paper, we use this notation to denote the polarizations $\mathcal{P}_{+}$ and $\mathcal{P}_{-}$ of light, respectively. To maintain the standard notation used in the literature for shifts, we have introduced the superscript $\mathcal{P}_{\pm}$ in Eq.~\eqref{redgeral1}. Thus, the subscript $\pm$ in this equation refers to the blueshift ($+$) and redshift ($-$), while $\mathcal{P}_{+}$ and $\mathcal{P}_{-}$ refer to the polarizations of light given by Eqs.~\eqref{eff_metric1} and~\eqref{eff_metric2}, respectively. From now on, we use this notation for the sake of simplicity. Furthermore, the kinematic shifts are obtained by noting that
\begin{equation}
\label{kinetic}\left(z_{\rm kin}^{\mathcal{P}_{\pm}}\right)_{\pm} = \left(z_{\rm tot}^{\mathcal{P}_{\pm}}\right)_{\pm} - \left(z_{\rm grav}\right)^{\mathcal{P}_{\pm}}.
\end{equation}
We also point out that the equations obtained in this section for the gravitational and kinematic shifts are consistent with the corresponding equations presented in Ref.~\cite{de2025exploring}.

\subsection{Main results}\label{subsec:mr}

The total redshift is a combination of the kinematic shift, also known as the Doppler shift, and the gravitational redshift. In short, the kinematic shift is driven by the relative motion between the radiation source and the observer. If the frequency decreases (increases), we have a redshift (blueshift). In turn, the gravitational redshift is due to the photon losing energy as it climbs out of a gravitational potential well. For simplicity, we investigate these effects separately, tracking the effects of the BH parameters and the impact of the effective geometry on both the kinematic and gravitational components of the total shift. Moreover, we restrict our analysis to asymptotic detectors, i.e., we assume that $\left(g_{tt}\right)_{d} \rightarrow -1$ as $r_{d} \rightarrow \infty$.

\begin{figure}[htbp]
\begin{centering}
    \includegraphics[width=\columnwidth]{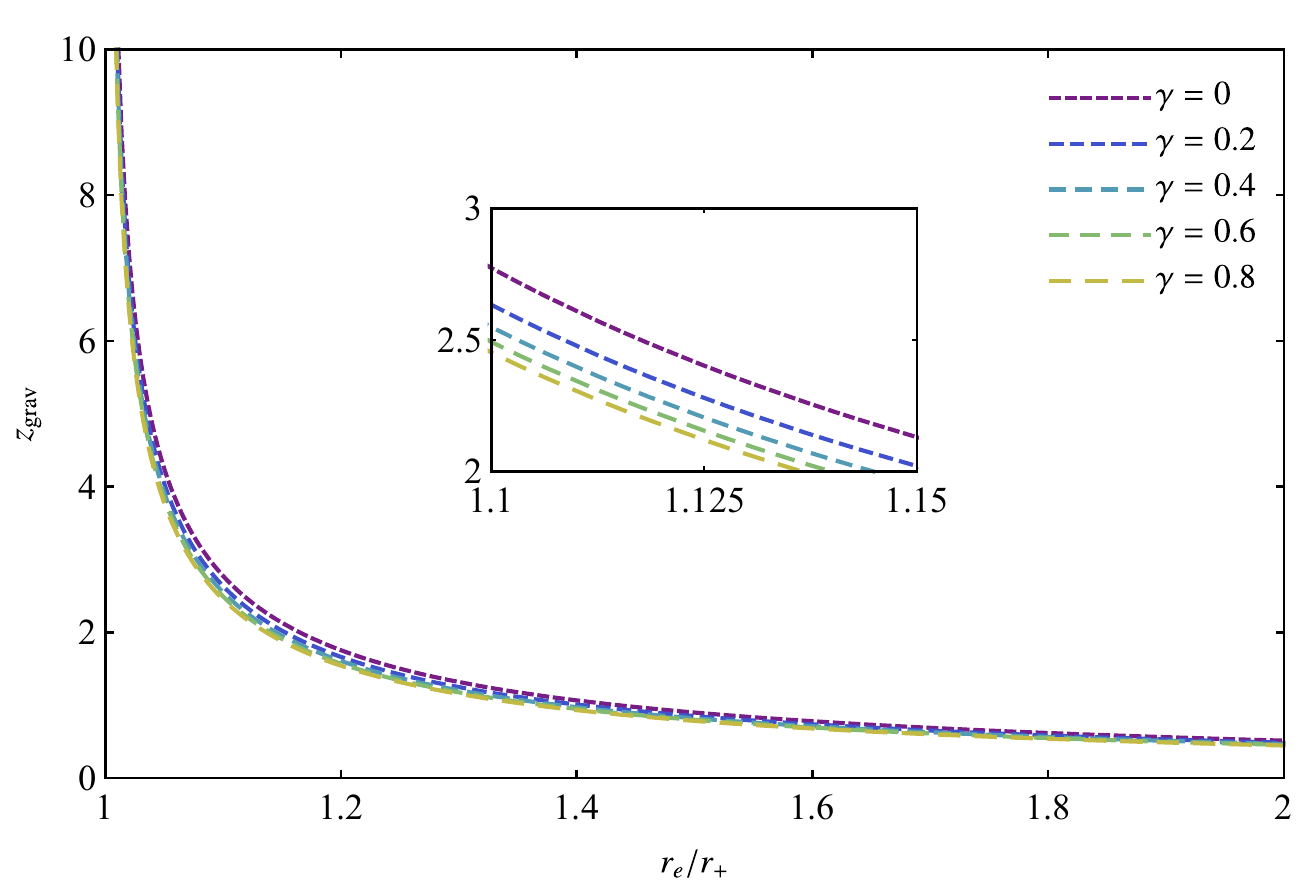}
    \caption{Gravitational redshift for the ModMax BH, as a function of $r_{e}/r_{+}$, considering distinct values of $\gamma$. The inset zooms the gravitational redshift near a given region. Here, we set $Q = 0.8M$.}
    \label{zgravfig}
\end{centering}
\end{figure}

Using Eq.~\eqref{zgrav2}, we can write the gravitational redshift for the ModMax BH, considering both polarizations of light, as
\begin{equation}
\label{zgrares}\left(z_{\rm{grav}}\right)_{+} = \left(z_{\rm{grav}}\right)_{-} = \dfrac{1}{\sqrt{f_{e}}} -1 \equiv z_{\rm{grav}} ,
\end{equation}
where $f(r_{e}) \equiv f_e$. For this case, notice that 
\begin{equation}
\lim_{r_{d} \rightarrow\infty}\left(\bar{g}_{tt}\right)_{+} = -e^{-2\gamma} \quad \text{and} \quad \lim_{r_{d} \rightarrow\infty}\left(\bar{g}_{tt}\right)_{-} = -1.
\end{equation}
Therefore, the gravitational redshift is insensitive to the birefringence of ModMax electrodynamics. This reinforces the findings for the Shapiro time delay in Appendix~\ref{appendix_invariance}. Just as a constant conformal rescaling of the $(t,r)$ sector leaves the physical time delay invariant, it also factors out entirely from the shift in the gravitational redshift measured by timelike observers. In Fig.~\ref{zgravfig}, we display the gravitational redshift for the ModMax BH. As we can see, the increase of $\gamma$ typically decreases the gravitational redshift. In other words, the photon loses less energy to overcome the gravitational potential well in the ModMax case than in the RN case. For an emitter far from the BH, the effects of $\gamma$ are negligible, as expected.

Using Eq.~\eqref{kinetic}, one can show that
\begin{equation}
\label{kinplus}\left(z_{\rm kin}^{\mathcal{P}_{+}}\right)_{\pm} = \sqrt{\dfrac{2}{2f_{e}-rf^{\prime}_{e}}}\left[1\mp \sqrt{\dfrac{e^{-2\gamma}rf^{\prime}_{e}}{2f_{e}}} \right]-\dfrac{1}{\sqrt{f_{e}}},
\end{equation}
for the polarization $\mathcal{P}_{+}$, while for $\mathcal{P}_{-}$, we get
\begin{equation}
\label{kinminus}\left(z_{\rm kin}^{\mathcal{P}_{-}}\right)_{\pm} = \sqrt{\dfrac{2}{2f_{e}-rf^{\prime}_{e}}}\left[1\mp \sqrt{\dfrac{rf^{\prime}_{e}}{2f_{e}}} \right]-\dfrac{1}{\sqrt{f_{e}}}.
\end{equation}
With Eqs.~\eqref{kinplus} and~\eqref{kinminus}, we can investigate how the BH parameters affect the kinematic shifts. Notice once more that the polarization of light $\mathcal{P}_{-}$ is conformally related to the spacetime metric [cf. Eq.~\eqref{eff_metric2}]. Therefore, we focus mainly on the results for the first polarization of light, since they are most influenced by the nonlinear electromagnetic fields.
\begin{figure*}[htbp]
\begin{centering}
    \includegraphics[width=\columnwidth]{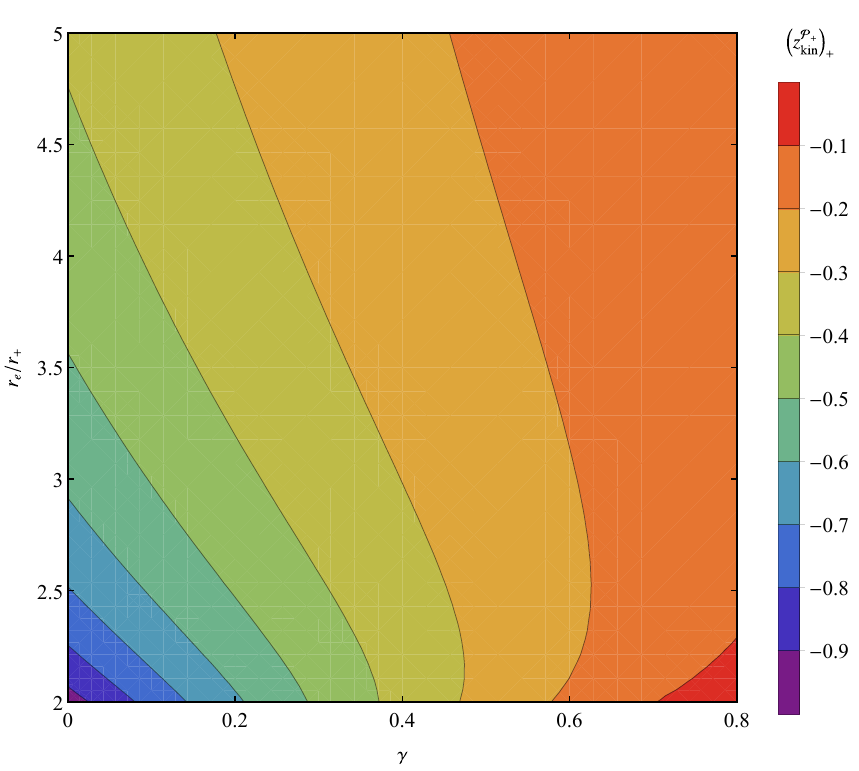}
    \includegraphics[width=0.985\columnwidth]{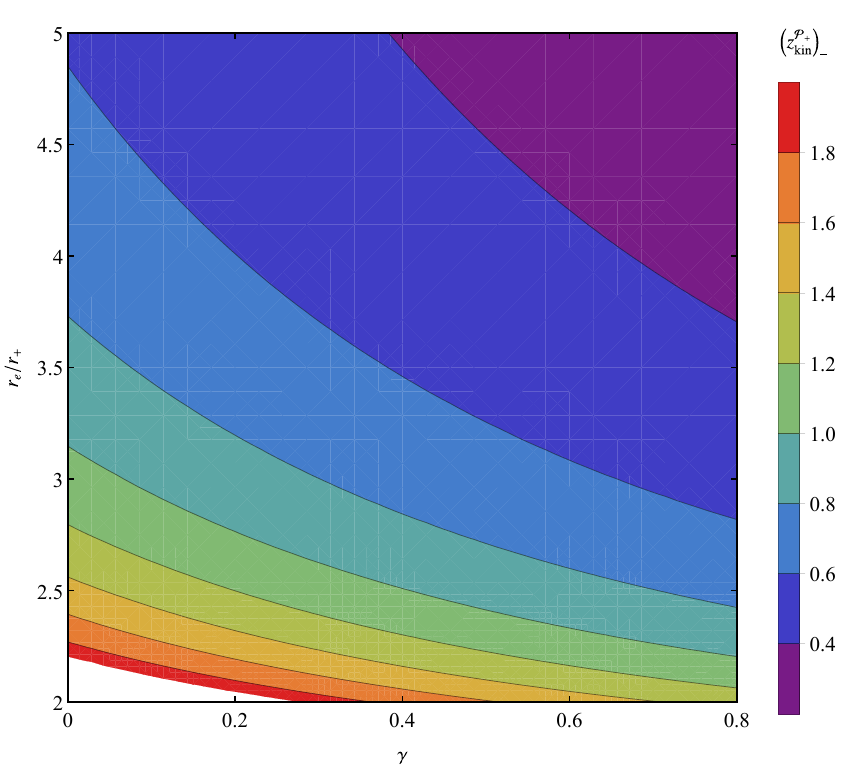}
    \caption{Blueshift (left panel) and redshift (right panel) for the ModMax BH, considering the polarization of light $\mathcal{P}_{+}$. We consider distinct values of $r_{e}/r_{+}$ and $\gamma$. Moreover, we set $Q = 0.8M$. Recall that, for this case, we assume asymptotic detectors ($r_{d} \rightarrow \infty$).}
    \label{boundgamma}
\end{centering}
\end{figure*}

In Fig.~\ref{boundgamma}, we show the blueshift and redshift, considering the polarization of light $\mathcal{P}_{+}$. As we can see, the blueshift decreases as we move away from the BH or increase $\gamma$. The contour lines indicate the configurations where the blueshift is constant. Since increasing $\gamma$ reduces the blueshift, the only way to maintain the same level of spectral shift for a larger $\gamma$ is to require that the photon be emitted from a region near the BH. Concerning the redshift, we observe that it decreases as we increase $r_{e}$ or $\gamma$. Moreover, the contour lines are interpreted in a similar way to the contour lines for the blueshift. We notice, however, that there is a white region for the  redshift case. This region is associated with the condition:
\begin{equation}
2f_{c}-r_{c}f^{\prime}_{c}=0,    
\end{equation}
which leads to
\begin{equation}
r_{c} = \dfrac{1}{2} \left(3M+ \sqrt{9M^{2}-8 e^{-\gamma}Q^2}\right).
\end{equation}
This equation provides the location of the null rings $r_{c}$ in the background of the spacetime metric. As the emitter cannot be placed within $r_{e} \leq r_{c}$, the white region in the figure associated with the redshift denotes these forbidden regions. Notice also that because the massive emitter cannot maintain a circular orbit within $r_e \leq r_c$, these regions are physically inaccessible for both kinematic shifts. To avoid ambiguity, we emphasize that we use the term ``decrease'' for the shifts in the sense of ``tending toward zero'', i.e., the effect approaches the limit of a flat spacetime.

\begin{figure}[htbp]
\begin{centering}
    \includegraphics[width=\columnwidth]{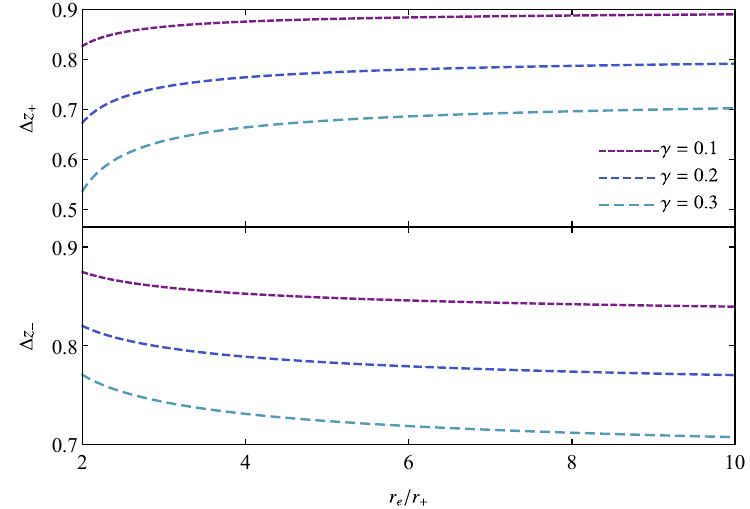}
    \caption{Ratio between the blueshift (top panel) and redshift (bottom panel), considering distinct values of $\gamma$ and $Q = 0.8M$. }
    \label{ratioredblue}
\end{centering}
\end{figure}

To quantify the degree of deviation from the results of the kinematic shifts between the effective geometries, we define
\begin{equation}
\label{ratio}\Delta z_{\pm} = \dfrac{\left(z_{\rm kin}^{\mathcal{P}_{+}}\right)_{\pm}}{\left(z_{\rm kin}^{\mathcal{P}_{-}}\right)_{\pm}}.
\end{equation}
This quantity measures the ratio between the kinematic blueshifts and redshifts for the first and second polarizations of light. In Fig.~\ref{ratioredblue}, we display the deviation rate given by Eq.~\eqref{ratio} for distinct values of $\gamma$. As we can see, the kinematic shifts for the first polarization of light are typically smaller than those for the second polarization of light, and this difference gets larger as we consider higher values of $\gamma$. We also notice that if the emitter is sufficiently close to the BH, we observe a sudden decrease (increase) in the deviation ratio for blueshift (redshift) results. Consequently, the kinematic shifts can be used to discriminate between the effective geometries.

\begin{figure}[htbp]
\begin{centering}
    \includegraphics[width=\columnwidth]{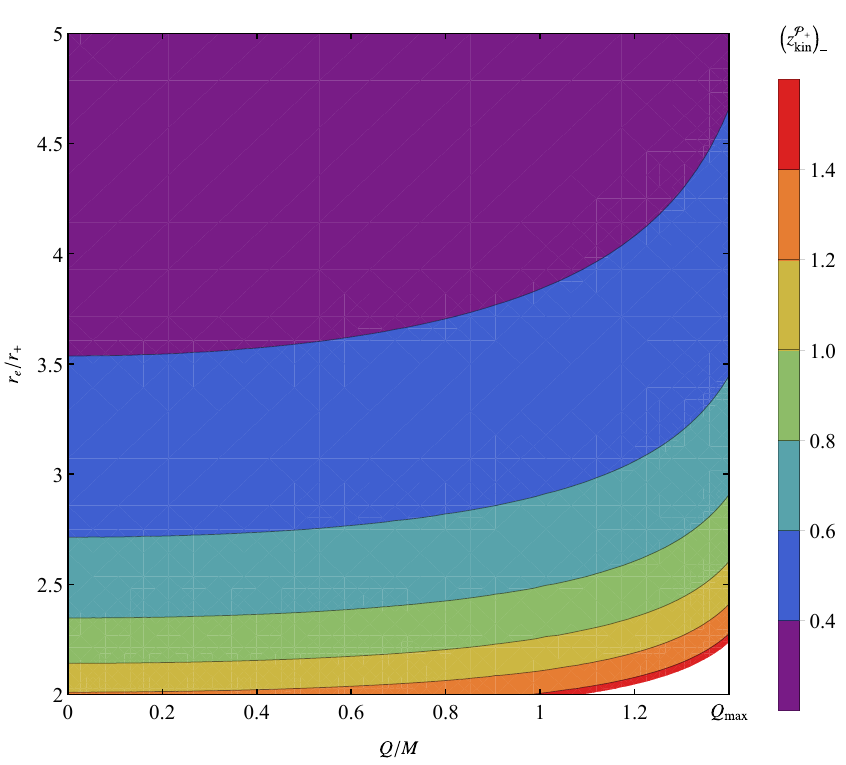}
    \caption{Redshift for the ModMax BH, considering the polarization of light $\mathcal{P}_{+}$, for distinct values of $r_{e}/r_{+}$ and $Q/M$. We set $\gamma = 0.8$ and $Q_{\rm{max}} \approx 1.4918M$ is the corresponding extremal charge.}
    \label{redshiftdiffQ}
\end{centering}
\end{figure}

For completeness, in Fig.~\ref{redshiftdiffQ}, we display the role of the BH charge in the redshift, considering the first polarization of light. As we can see, the redshift decreases as we move away from the BH, but it increases as we consider higher values of $Q/M$. Here, we also observe the forbidden regions (white area)  for moderate-to-extremal BH charge-to-mass ratios. 

\section{Massive Particle Effects}\label{sec:mpe}

In the previous sections, we analyzed the propagation of light in the effective geometries associated with ModMax electrodynamics. The background geometry controls the dynamics of massive test particles and stationary observers. This section explores classical relativistic effects encoded directly in the ModMax BH metric, considering  the orbital precession of massive particles. In particular, these observables will be used to place constraints on the parameter $\gamma$, which characterizes the nonlinear nature of the ModMax electrodynamics, complementing our previous investigations.

We now compare the orbital periapsis precession of the S2 star around Sgr A$^{\star}$, as measured by the GRAVITY Collaboration \cite{GRAVITY:2020gka} and interpreted within the framework of the Schwarzschild spacetime. The primary objective of this analysis is to investigate whether meaningful constraints can be placed on the parameter $\gamma$, which characterizes the nonlinear nature of the ModMax solution. The motion of a test particle along timelike geodesics is subject to the condition
\begin{equation}
g_{\mu\nu}\dot{x}^{\mu}\dot{x}^{\nu}=-1\,,
\label{gmn}
\end{equation}
and Eq.~\eqref{AB} now becomes
\begin{eqnarray}
    -A(r)\dot{t}^2+B(r)\dot{r}^2+r^2\dot{\varphi}^2=-1\,,
    \label{eq:dsmov}
\end{eqnarray}
from which, using Eqs.~\eqref{moment} and \eqref{energy}, we obtain
\begin{eqnarray}
    \frac{d\varphi}{dr}=\frac{\sqrt{A(r)B(r)}}{r^2}L\left[E^2-A(r)\left(1+\frac{L^2}{r^2}\right)\right]^{-1/2}\,.\label{dvarphi}
\end{eqnarray}

The orbital precession is given by \cite{zhou2020precessing, zhang2022probing, lin2023precessing}
\begin{equation}
\delta\omega=\delta\varphi-2\pi \,, 
\label{deltaomega}
\end{equation} 
where $\delta\varphi$ is the total azimuthal angle swept out during one orbital period in the equatorial plane, given by
\begin{equation}
\delta\varphi=2\int^{r_a}_{r_p}\frac{d\varphi}{dr}dr=2\int^{\pi}_{0}\frac{d\varphi}{d\chi}d\chi \,,
\label{intdelta}
\end{equation} 
where $r_a=a(1+\epsilon)$ and $r_p=a(1-\epsilon)$ are the apoastron and periastron, respectively, depending on the semi-major axis $a$ and the eccentricity $\epsilon$ of the orbit. To evaluate the integral, we parameterize the radial coordinate in terms of the relativistic anomaly $\chi$ as
\begin{equation}
r=\frac{a(1-\epsilon^2)}{1+\epsilon\cos\chi}\,,\label{cord1}
\end{equation}
where $r_p$ corresponds to $\chi=0$ and $r_a$ to $\chi=\pi$, respectively.
By using $\frac{d\varphi}{d\chi}=\frac{d\varphi}{dr}\frac{dr}{d\chi}$ with \eqref{cord1}, we can rewrite Eq.\,\eqref{dvarphi} as
\begin{eqnarray}
\frac{d\varphi}{d\chi}&=&\frac{\sqrt{A(\chi)B(\chi)}}{r^2(\chi)}a\,\epsilon\, L\frac{\left(1-\epsilon^2\right)\sin\chi}{\left(1+\epsilon\cos\chi\right)^2}\nonumber\\
&&\times\left[E^2-A(\chi)\left(1+\frac{L^2}{r^2(\chi)}\right)\right]^{-1/2}\,. \label{2d}
\end{eqnarray}
Since the radial velocity vanishes at the turning points, namely  $\dot{r}^2\big|_{r_{a,p}}=0$, the angular momentum and energy can be expressed as
\begin{equation}
L^2=\frac{r^2_a\,r^2_p\left[A(r_p)-A(r_a)\right]}{r^2_pA(r_a)-r^2_aA(r_p)}\,,
\label{L}
\end{equation}
and
\begin{equation}
E^2=\frac{A(r_p)A(r_a)\left(r^2_p-r^2_a\right)}{A(r_a)r^2_p-A(r_p)r^2_a}\,.
\label{E}
\end{equation}
Now, let us consider that the Sgr A$^{\star}$ is described by the metric function \eqref{MF_EH}. Inserting Eqs. \eqref{L} and \eqref{E} into Eq. \eqref{2d}, considering \eqref{cord1} and integrating by \eqref{intdelta}, we get Eq. \eqref{deltaomega}. Treating the effects of $\gamma$ as a perturbation, we obtain the following expression for the orbital precession:
\begin{eqnarray}
  \delta\omega_{\rm MM} =\frac{2\pi M}{a(1-\epsilon^2)}
\left(3 - 2e^{-\gamma}q^2\right)\,, \label{omegaMM}
\end{eqnarray} 
where we introduce the dimensionless charge $q \equiv Q/2M$. As expected, the RN result is recovered in the limit $\gamma \rightarrow 0$ \cite{zhang2022probing}, and the GR result is recovered in the limit $q \rightarrow 0$, i.e., 
\begin{eqnarray}
    \delta\omega_{\rm{GR}}=\frac{6M\pi}{a(1-\epsilon^2)}\,.\label{omegaSC}
\end{eqnarray}

By taking the ratio between the orbital precessions \eqref{omegaMM} and \eqref{omegaSC}, we find that
\begin{equation}
f=\frac{\delta\omega_{\rm{MM}}}{\delta\omega_{\rm{GR}}}=1-\frac{2 e^{-\gamma } q^2}{3 }\,,\label{f}
\end{equation}
where in the limit $ q\rightarrow 0$, we have $f \rightarrow 1$, as expected. Let us now assume values determined by the GRAVITY collaboration \cite{GRAVITY:2020gka} for the orbital precession of the star $S2$ around Sgr A$^{\star}$.  The parameters $\epsilon=0.885$ and $a=1031$ are taken into account, as well as the mass $M_{\rm{Keck}}= 3.951 \times 10^6M_\odot$ \cite{do2019relativistic}, for which the values for $f$ provided by \cite{GRAVITY:2020gka} are $f=1.10 \pm 0.19$. Using the observational result in  Eq. \eqref{f} and defining $\beta=e^{-\gamma}q^2$, the above inequality becomes $0.91 \leq 1-\frac{2}{3}\beta \leq 1.29$. Solving the two inequalities separately yields $\beta \leq 0.135$ and $\beta \geq -0.435$.

\begin{figure}[htbp]
\begin{centering}
    \includegraphics[width=\columnwidth]{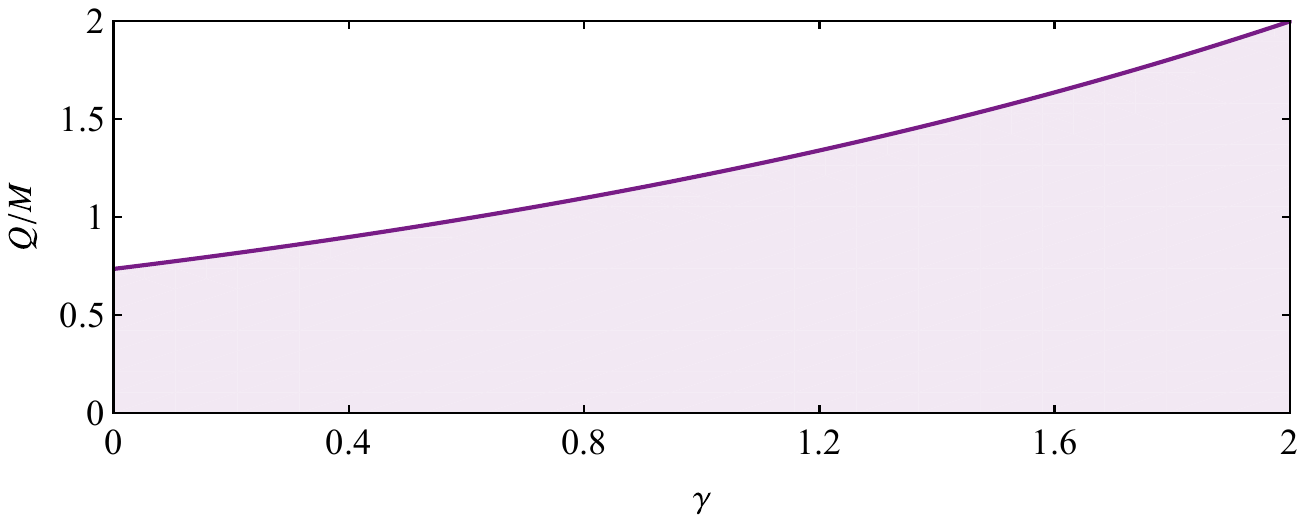}
    \caption{Constraint given by Eq.~\eqref{constraint} for some values of $\gamma$ and $Q/M$. The solid purple line correspond to the saturation of the upper bound, namely, $e^{-\gamma}\left(Q/M\right)^2 = 0.54$.}
    \label{boundgamma2}
\end{centering}
\end{figure}

However, since $\beta=e^{-\gamma}q^2$, with $e^{-\gamma}>0$ and $q^2\geq0$, one necessarily has $\beta\geq0$. Therefore, the lower bound has no physical meaning, and the observational constraint leads to
\begin{equation}
\label{constraint}0 \leq e^{-\gamma}q^2 \leq 0.135.    
\end{equation}
For completeness, in Fig.~\ref{boundgamma2}, we display this constraint for some values of $\gamma$ and $Q/M$. As we can see, the values of $Q/M$ depend heavily on the chosen values of $\gamma$. This is consistent with the event horizon structure of the ModMax BH [cf Eq.~\eqref{extcase}], in which the extreme charge values depend on $\gamma$. Remarkably, this result shows that the orbital precession of the S2 star constrains only the combination $ e^{-\gamma}q^2$, rather than the ModMax parameter $\gamma $ itself. Consequently, the observational data alone cannot break the degeneracy between the nonlinear parameter and the BH charge.

A more restrictive condition emerges when the observational constraint is combined with the existence of an event horizon. For ModMax BH, the horizon condition requires $0\leq Q\leq Me^{\gamma/2} $. Writing the charge as $Q=\alpha e^{\gamma/2}M$, with $(0\leq \alpha \leq 1)$, where $\alpha=1$ corresponds to the extremal configuration, the observational bound becomes simply $\alpha^2 \leq C$. For the S2 data, this yields $\alpha \lesssim 0.73$. Therefore,
\begin{equation}
Q \lesssim 0.73 \,e^{\gamma/2}M =0.73\,Q_{\rm ext},
\end{equation}
showing that the allowed charge must remain safely below its extremal value. Notably, the dependence on $ \gamma $ disappears from the final inequality. Hence, although the orbital precession data do not directly constrain the $\gamma$ parameter, they exclude highly charged and near-extremal ModMax BHs as viable candidates for Sgr A$^{\star}$. In other words, any ModMax BH compatible with the S2 observations must possess a charge smaller than approximately  $73 \%$ of the extremal charge, independently of the value of $\gamma$.

\section{Concluding remarks}\label{sec:cr}

Phenomenologically, one of the most fascinating outcomes of NED theory is that photons propagate along null geodesics governed by an effective geometry. In particular, when considering a two-parameter electromagnetic Lagrangian density, as is the case with ModMax electrodynamics [cf. Eq.~\eqref{modmax}], vacuum birefringence can occur. In this context, photons may follow null geodesics according to two distinct effective geometries [cf. Eq.~\eqref{eff_metric}]. Here, we aimed to explore signatures of nonlinear electromagnetic fields and constrain the additional parameters of NED-sourced BH geometries based on some GR experimental tests. To do this, we investigated the ModMax BH geometry via light and orbital mechanics, considering the Shapiro time, the Sagnac effect, the gravitational and kinematic shifts, and the orbital precession.

The results obtained in Sec.\,\ref{Shap} show that both effective metrics yield exactly the same photon propagation time and, consequently, the same Shapiro time delay. A similar feature has been reported previously for the RN and Ayón-Beato-García BH spacetimes, whose analytical expressions for the photon travel time are identical, making these solutions indistinguishable through the Shapiro effect alone \cite{Junior:2023nku}. It is worth emphasizing, however, that, unlike Ref. \cite{Junior:2023nku}, where the comparison is performed between the spacetime metrics of two distinct BH solutions, our analysis is based on the effective metrics governing photon propagation in ModMax electrodynamics. Therefore, the equality found here is not a consequence of comparing different spacetime geometries, but rather of the effective description of light propagation associated with the birefringent structure of the theory. As demonstrated in subsection \ref{subsec:tdem}  and formally proved in Appendix \ref{appendix_invariance}, this result follows from an invariance property of the integral under a constant rescaling of the temporal and radial sectors of the effective metrics.

The numerical analysis confirms the analytical results and shows that the ModMax corrections remain extremely small throughout the weak field regime. For both stellar-mass and supermassive BHs, the relative deviations with respect to the RN prediction remain at the level of $10^{-4}$ over the range of the parameter $\gamma$ considered. As predicted by the analytical expression, increasing $\gamma$ progressively suppresses the charge dependent contribution, producing only a slight increase in the total Shapiro time delay while leaving the dominant mass dependent logarithmic term unchanged. Consequently, although the effective metrics arise from distinct polarization modes of photon propagation, the Shapiro effect is insensitive to birefringence and cannot distinguish between the two effective geometries. Moreover, the small deviations with respect to the Maxwell limit indicate that, in the weak field regime, the Shapiro time delay is only weakly affected by the NED corrections introduced by ModMax electrodynamics. These findings suggest that observable signatures of the effective geometry should instead be sought in null geodesic phenomena that are intrinsically more sensitive to the birefringent nature of the theory.

In this sense, the Sagnac effect also provides an interesting mechanism to investigate potential effects of NED fields. We have seen that for ModMax BHs with $\gamma > 10^{-11}$, one could theoretically measure the relative deviation in the Sagnac effect between different light polarizations, assuming identical BH parameters. This finding implies that the Sagnac effect could serve as a practical tool to detect vacuum birefringence in gravitational environments. As discussed earlier, the primary prerequisites are an astrophysical body with an electromagnetic field strong enough to induce NED effects, coupled with instruments capable of detecting the Sagnac effect. Such a scenario is particularly relevant for magnetars, given their intense magnetic fields. Moreover, the BH charge-to-mass ratio suppresses the Sagnac effect, whereas the NED parameter contributes to its enhancement. We also found that when comparing the results obtained in ModMax with those obtained in RN, even minimal values of $\gamma$ can produce substantial differences if the sources are sufficiently fast. As an example, for the light polarization $\mathcal{P}_{+}$, setting $\gamma = 0.01$ and $\bar{\Omega}_{0} = 0.95$ yields a deviation of approximately $12\%$ with respect to the RN case, assuming the same BH charge-to-mass ratio.

Concerning the gravitational and kinematic shifts, we have seen that the gravitational redshift is also insensitive to the birefringence structure of the ModMax electrodynamics. Moreover, both the kinematic blueshift and redshift decrease as we increase $r_{e}$ or $\gamma$, and the only way to maintain the same level of spectral shift for a larger $\gamma$ is to bring $r_{e}$ closer and closer to the BH. We have also compared the ratio of the kinematic shifts between the two polarizations of light. We found that the kinematic shifts for the first polarization of light are typically smaller than those for the second polarization of light. In addition to that, the difference gets larger as we consider higher values of $\gamma$. This shows that the kinematic shifts can be used to distinguish between both polarizations of light. For completeness, we investigated the role of the BH charge in the redshift. We found that the redshift decreases as we move away from the BH, but it increases as we consider higher values of $Q/M$. We also observed that it is possible to find forbidden regions (white areas) in the spectral lines of the shifts. These regions are related to the location of the emitter, which must be larger than the null ring location.

In Sec.\,\ref{sec:mpe}, we have studied the orbital precession of the S2 star in the ModMax BH spacetime \eqref{MF_EH}. The resulting correction depends only on the combination $e^{-\gamma}q^2$, leading to a degeneracy between the NED parameter and the BH charge. Using the GRAVITY Collaboration \cite{GRAVITY:2020gka} measurement of the S2 precession, we obtain the bound $e^{-\gamma}q^2 \leq 0.135$. This constraint does not translate into an independent limit on $\gamma$. However, when combined with the horizon condition, it implies $Q \lesssim 0.73Q_{\rm ext}$, independently of $\gamma$. Thus, highly charged and near-extremal ModMax BHs are disfavored by the S2 data.

We also point out that our work provides a step-by-step procedure for exploring the signatures of NED fields and constraining the additional free parameters of the corresponding NED-sourced BH models in light of some GR experimental tests. Some interesting results can be anticipated from a spherically symmetric setup, as we have obtained, but generalizations considering rotating spacetimes are necessary due to their astrophysical relevance. In addition, we considered the ModMax geometry, but there are other well-motivated NED-based BH solutions. Notable examples are the Euler-Heisenberg~\cite{Ruffini:2013hia,Yajima:2000kw} and Born-Infeld~\cite{Cai:2004eh,Myung:2008eb} BHs. Both have a two-parameter electromagnetic Lagrangian density with one free parameter associated with the corresponding NED theory. In particular, their corresponding effective geometries have a structure that is very different from those of ModMax, which could potentially break the degeneracy that occurs in the Shapiro time delay and gravitational redshift for the ModMax case. Finally, we emphasize that it would also be interesting to investigate the experimental tests discussed here in NED-based BH solutions obtained within the framework of alternative theories of gravity~\cite{Yang:2020jno,Sekhmani:2025epe,Ali:2022yys}.

We conclude this paper by noting that we are not claiming here that the phenomenon of vacuum birefringence can be measured based on the ModMax geometry, nor that this model is the best one for exploring the effects of NED fields. The aim of this paper is purely theoretical in the sense of quantifying the extent to which the effective geometry (in our particular case, vacuum birefringence as well) plays a role within the context of a well-motivated NED model. Therefore, our goal was twofold. First, to investigate which potential scenarios the effects of effective geometries could alter. Second, to explore how these scenarios could be used to better understand the role of the effective geometries and the NED fields as well.

\begin{acknowledgments}

We acknowledge the Conselho Nacional de Desenvolvimento Cient\'ifico e Tecnol\'ogico (CNPq) from Brazil for partial financial support. M. A. A. de Paula is supported by CNPq/PDJ 150589/2025-5, and E. L. B. Junior is supported by CNPq/PQ 307085/2026-0. M. A. A. de Paula would like to thank Haroldo Cilas Duarte Lima Júnior and Mustapha Azreg-A\"{\i}nou for some insightful discussions.

\end{acknowledgments}

\appendix

\section{Energy and angular momentum of the photon}\label{breton}

As discussed in the Introduction, the authors of Ref.~\cite{Guzman-Herrera:2023zsv} also investigated the gravitational and kinematic shifts in the background of the ModMax BH geometry. In this appendix, we detail why our results differ from their calculated results and why ours are more consistent with the literature.

The nonlinearities of the electromagnetic field affect the motion of photons in such a way that light propagates according to an effective light cone. Moreover, if the corresponding NED model depends on two electromagnetic scalars, we may have vacuum birefringence. This is the case of the ModMax BH [cf. Eqs.~\eqref{eff_metric1}-\eqref{eff_metric2}], and also of the Euler-Heisenberg BH (see, e.g., Ref.~\cite{dePaula:2026blu}). As the corresponding geometry where photons follow null geodesics is modified, i.e., it does not correspond to the spacetime metric [see, e.g., Eq.~\eqref{LE}], the definition of energy and angular momentum also changes.

In NED-sourced spacetimes, the energy and angular momentum of the photon are conserved quantities with respect to the Lagrangian (or, equivalently, the Hamiltonian) related to the corresponding effective geometry. This is the view taken by most investigations that consider the effective geometry in NED-based BH spacetimes (see, e.g., Refs.~\cite{Toshmatov:2021fgm,Stuchlik:2019uvf,Allahyari:2019jqz,Kruglov:2020tes,Liang:2017vdd,Liang:2017wym,dePaula:2023ozi,dePaula:2026blu,dePaula:2024yzy,Murk:2024nod} and references therein).

The authors of Ref.~\cite{Guzman-Herrera:2023zsv} defined the energy and angular momentum of a test particle as [cf. their Eq. (4.14)]
\begin{equation}
\label{a1}E = g_{tt}\dfrac{dt}{d\tau} \quad \text{and} \quad L = g_{\varphi \varphi}\dfrac{d\varphi}{d\tau},
\end{equation}
respectively, where $\tau$ is an affine parameter. For convenience, we keep the metric components general (see the footnote~\ref{nota}). They consider the correct equations for the effective metric of photons in the ModMax geometry [cf. their Eqs.~(3.4) and~(3.5)]. In our notation, the effective metrics are given by Eqs.~\eqref{eff_metric1} and~\eqref{eff_metric2}, respectively. However, the authors used the incorrect definition of energy and angular momentum for the photon in NED. More specifically, they consider that these conserved quantities satisfy Eq.~\eqref{a1}, rather than
\begin{equation}
\label{correcta1}\left(\bar{E}\right)_{\pm} = -\left(\bar{g}_{tt}\right)_{\pm}\dfrac{dt}{d\tau} \quad \text{and} \quad \left(\bar{L}\right)_{\pm} = \left(\bar{g}_{\varphi \varphi}\right)_{\pm}\dfrac{d\varphi}{d\tau}.
\end{equation}
The indices $\pm$ denote the energy and angular momentum of photons in the first and second polarizations of light, respectively. This is further emphasized when the authors investigate the gravitational and kinematic redshifts of the ModMax BH. They use the eikonal equation for photons in NED, i.e., 
\begin{equation}
\left(\bar{g}_{\mu\nu}\right)_{\pm}k^{\mu}k^{\nu} = 0,
\end{equation}
where $k^{\mu}$ are null vectors. In light of the arguments presented above, $k^{\mu}$ should be null vectors with respect to the effective geometries, considering each polarization of light. However, the authors used $k^{t} = -E/f(r)$ and $k^{\varphi} = L/r^{2}$ (see, e.g., Sec. 4.4 of Ref.~\cite{Guzman-Herrera:2023zsv}, in particular, Eq. (4.49) and the following discussions), which are null vectors with respect to the standard geometry [cf. Eq.~\eqref{a1}]. This is conceptually incorrect because null vectors computed in the background of the effective geometry are not necessarily null with respect to the standard geometry~\cite{dePaula:2023ozi,dePaula:2024yzy}. Since the second polarization of light in ModMax electrodynamics is conformally related to the standard geometry [cf. Eq.~\eqref{eff_metric2}], this ``misuse of notation'' does not affect the results. However, the first polarization of light is not conformally related to the spacetime metric [cf. Eq.~\eqref{eff_metric1}], which does affect the results, as we detail below.

This incorrect definition of the energy and angular momentum of the photon used in Ref.~\cite{Guzman-Herrera:2023zsv} affects all physical quantities/observables associated with the motion of photons. In particular, for the polarization $\mathcal{P}_{+}$ [see Eq. (11) of Ref.~\cite{Guzman-Herrera:2023zsv}], the authors inevitably lose a factor of $e^{-2\gamma}$ during algebraic manipulations since $\left(\bar{g}_{tt}\right)_{+} = e^{-2\gamma}g_{tt}$ and $\left(\bar{g}_{rr}\right)_{+} = e^{-2\gamma}g_{rr}$. Here, for the sake of simplicity and given the primary focus of our paper, we focus solely on resolving this issue considering the gravitational and kinematic redshifts.

\section{Proof of the invariance of the Shapiro time delay}\label{appendix_invariance}

In Sec.~\ref{Shap}, we notice that the Shapiro time delay is exactly the same for both effective metrics arising in ModMax electrodynamics, despite the fact that they correspond to different effective geometries associated with the two polarization modes. In this appendix, we show that this result is not accidental but follows from the particular conformal relation between the two metrics in the ($t,r$) sector of the metrics.

Indeed, the first effective metric is defined by
\begin{equation}
A_{1}(r)=e^{-2\gamma}f(r), \qquad
B_{1}(r)=\frac{e^{-2\gamma}}{f(r)}
\end{equation}
whereas the second one is
\begin{equation}
A_{2}(r)=f(r), \qquad
B_{2}(r)=\frac{1}{f(r)},
\end{equation}
Therefore,
\begin{equation}\label{A1B1}
A_{1}(r)=e^{-2\gamma}A_{2}(r),
\qquad
B_{1}(r)=e^{-2\gamma}B_{2}(r),
\end{equation}
showing that both effective metrics differ only by a constant conformal factor in the $(t,r)$ sector of the spacetime metrics.

The propagation time is obtained from
\begin{equation}
\label{progtime}\frac{dt}{dr}= \frac{1}{b}
\left[ \frac{A(r)}{B(r)}
\left( \frac{1}{b^{2}}
- \frac{A(r)}{C(r)}
\right) \right]^{-1/2},
\end{equation}
where the impact parameter is defined as
\begin{equation}
b^{2}=\frac{C(d)}{A(d)}.
\end{equation}
Under the above conformal rescaling, one finds
\begin{equation}
b_{1}^{2} = \frac{C(d)}{A_{1}(d)} =
e^{2\gamma}b_{2}^{2}.
\end{equation}
Therefore,
\begin{equation}
b_{1}=e^{\gamma}b_{2},
\end{equation}
while
\begin{equation}
\frac{A_{1}}{B_{1}} = \frac{A_{2}}{B_{2}} = f^{2}(r).
\end{equation}
Furthermore,
\begin{equation}
\frac{1}{b_{1}^{2}} - \frac{A_{1}}{C} = e^{-2\gamma} \left( \frac{1}{b_{2}^{2}} - \frac{A_{2}}{C}\right).
\end{equation}
Substituting these relations into Eq.~\eqref{progtime}, we get
\begin{equation}
\begin{split}
\left(\frac{dt}{dr}\right)_{1}
&=
\frac{1}{e^{\gamma}b_{2}}
\left[ f^{2}(r)\, e^{-2\gamma}
\left( \frac{1}{b_{2}^{2}} - \frac{A_{2}}{C} \right) \right]^{-1/2}
\\
&=
\frac{1}{e^{\gamma}b_{2}} \,e^{\gamma} \left[ f^{2}(r) \left( \frac{1}{b_{2}^{2}}
- \frac{A_{2}}{C} \right) \right]^{-1/2}\\
&=
\left(\frac{dt}{dr}\right)_{2},
\end{split}
\end{equation}
where the conformal factors cancel exactly. Consequently, the integrands corresponding to the two effective metrics are identical, implying 
\begin{equation}
T_{1}=T_{2}, \qquad
\Delta T_{1}=\Delta T_{2}.
\end{equation}

Therefore, the equality of the Shapiro time delay does not result from the weak field expansion, but rather from the fact that the two effective metrics are related by a constant conformal transformation. Such a transformation leaves the null structure of the spacetime unchanged and produces an exact cancellation of the conformal factor in the propagation integral. Hence, although the two effective metrics describe different polarization modes in ModMax electrodynamics, they predict the same Shapiro time delay within the geometric optics approximation.

\section{Conversion from units}\label{app:units}

The analytical expressions are derived in geometrized units, adopting $G=c=1$. Accordingly, the mass parameter is expressed in units of the solar mass, $M=mM_\odot$, and the resulting time delays are obtained in the corresponding geometrized units. In the numerical evaluation, the mass dependence is retained explicitly through $m$, while the conversion to physical time units is performed at the final stage using
\begin{eqnarray}
1 M_\odot \rightarrow \frac{GM_\odot}{c^3}
=4.92549\times10^{-6}\, {\rm s}.
\end{eqnarray}
Hence, the values displayed in Tables \ref{tab:stellar} and \ref{tab:sgrA} in seconds are obtained by multiplying the geometrized results by $GM_\odot/c^3$. For example, for $M=6.1M_\odot$, the geometrized result $\Delta T_{\rm RN}=524.459$ corresponds to
\begin{eqnarray}
    \Delta T_{\rm RN}
=524.459\frac{GM_\odot}{c^3}
\simeq 2.583\times10^{-3}\ {\rm s}.
\end{eqnarray}
This conversion restores the physical time units while leaving the relative differences between the ModMax and RN predictions unchanged.

\bibliography{ref}

\end{document}